\documentclass[a4paper,11pt]{article}
\pdfoutput=1

\usepackage[english]{babel}
\usepackage{jheppub}
\usepackage[T1]{fontenc}
\usepackage{graphicx,calc}
\usepackage{epsfig}
\usepackage{amsmath}
\usepackage{amssymb}
\usepackage{cleveref}
\usepackage{float}
\usepackage{placeins}
\usepackage{braket}
\usepackage{ifthen}
\usepackage{slashed}
\usepackage{mathdots}
\usepackage{feynmf}
\usepackage{color}
\usepackage{multicol}
\usepackage{caption}
\usepackage{subcaption}
\usepackage{array}
\usepackage[export]{adjustbox}
\usepackage{subcaption}
\usepackage{multicol}
\usepackage{comment}
\usepackage{colortbl}
\usepackage{xcolor}
\usepackage{booktabs}

\newcommand{\eps}{\varepsilon}
\newcommand{\ii}{{\mathrm{i}}}
\newcommand{\dd}{{\mathrm{d}}}

\newcommand{\dlog}{\dd \log}

\newcommand{\PBA}{\ensuremath{\mathrm{PB}_A}}
\newcommand{\PBB}{\ensuremath{\mathrm{PB}_B}}
\newcommand{\PBC}{\ensuremath{\mathrm{PB}_C}}
\newcommand{\PBD}{\ensuremath{\mathrm{PB}_D}}
\newcommand{\PBE}{\ensuremath{\mathrm{PB}_E}}
\newcommand{\PBF}{\ensuremath{\mathrm{PB}_F}}

\newcommand{\MISectorGroup}[1]{\par\addvspace{2\bigskipamount}
\noindent\makebox[\linewidth]{\textbf{#1}}\par\nobreak\medskip}
\newcommand{\MISector}[3]{\par\addvspace{\bigskipamount}
\noindent
\begin{minipage}{\linewidth}
\textbf{Sector $#1$}\par\medskip
\begin{minipage}[c]{0.5\linewidth}
\centering #2
\end{minipage}%
\begin{minipage}[c]{0.5\linewidth}
\centering $\displaystyle #3$
\end{minipage}
\end{minipage}\par}
\allowdisplaybreaks

\title{Two-loop integrals for $t\bar{t}\gamma$ production at hadron colliders in the leading-colour approximation}

\author[a]{Souvik Bera,}
\author[b,c]{Colomba Brancaccio,}
\author[d,e]{Dhimiter Canko,}
\author[e]{Federico Ripani,}
\author[f]{Simone Zoia}

\affiliation[a]{Asia Pacific Center for Theoretical Physics, Pohang, 37673, Korea}
\affiliation[b]{Max-Planck-Institut für Physik, Boltzmannstr. 8, 85748 Garching, Germany}
\affiliation[c]{Max Planck-IAS-NTU Center for Particle Physics, Cosmology and Geometry}
\affiliation[d]{Dipartimento di Fisica e Astronomia, Università di Bologna \&
INFN, Sezione di Bologna, via Irnerio 46, 40126 Bologna, Italy}
\affiliation[e]{Dipartimento di Fisica and Arnold-Regge Center, Università di Torino \& INFN, Sezione di Torino, Via P. Giuria 1, I-10125 Torino, Italy}
\affiliation[f]{INFN, Sezione di Milano, Via Celoria 16, 20133 Milano, Italy}

\emailAdd{souvik.bera@apctp.org}
\emailAdd{colomba@mpp.mpg.de}
\emailAdd{dhimiter.canko@unito.it}
\emailAdd{federico.ripani@unito.it}
\emailAdd{simone.zoia@mi.infn.it}

\preprint{MPP-2026-197, TIF-UNIMI-2026-14}

\abstract{
We complete the calculation of the two-loop Feynman integrals required for the next-to-next-to-leading order QCD corrections to on-shell top--antitop production in association with a photon at hadron colliders, in the leading-colour approximation.
The same integrals also enter the subleading-colour corrections to top--antitop production in association with a jet.
We construct integral bases that satisfy algebraic differential equations. 
These are largely in canonical form; the exceptions are a few entries, associated with elliptic curves or nested square roots, that depend at most quadratically on the dimensional regulator $\eps$.
We solve the differential equations in terms of a tailored set of special functions, which will enable the analytic extraction of the two-loop finite remainder.
We evaluate these functions by numerically solving the differential equations they satisfy, and show that the resulting routine is fast and stable enough in the physical region for phenomenological applications.
}

\begin{document}
\maketitle
\flushbottom

\section{Introduction}
\label{Introduction}

Top-quark production processes play a prominent role in the precision programme of the Large Hadron Collider (LHC). 
Among these, top-quark pair production in association with a photon ($t\bar{t}\gamma$) offers direct access to the electromagnetic properties of the top quark and therefore constitutes a powerful probe of the Standard Model (SM) as well as of possible new-physics effects. 
Indeed, $t\bar{t}\gamma$ production enables a direct measurement of the top-quark electric charge~\cite{Baur:2001si} and probes the structure of the top--photon coupling~\cite{Baur:2004uw}. Furthermore, any deviation of the measurements from the SM prediction could be a hint of a new exotic top-like particle or of an anomalous electric dipole moment of the top quark~\cite{Fael:2013ira,Bouzas:2012av,Aguilar-Saavedra:2014vta,Schulze:2016qas,Etesami:2016rwu}, as studied in the context of effective field theories~\cite{BessidskaiaBylund:2016jvp}. 
Although the cross section for $t\bar{t}\gamma$ production at the LHC is much smaller than that for top-quark pair production alone, the growing amount of data has enabled the ATLAS and CMS collaborations to carry out extensive measurements of both the inclusive cross section and differential distributions for this process, with recent results published in refs.~\cite{ATLAS:2024hmk,CMS:2025zbe}.
As the precision of experimental measurements at the LHC increases, more accurate theoretical predictions for this process are required~\cite{Huss:2025nlt}.

Fixed-order NLO QCD and EW predictions have so far been computed~\cite{Duan:2009kcr,Duan:2011zxb,Duan:2016qlc,Maltoni:2015ena}, and the so-called complete NLO predictions are also available~\cite{Pagani:2021iwa}. 
Considerable effort has gone into a more realistic description of this process, taking into account the effects of top-quark decays. NLO QCD theoretical predictions have been combined with a parton shower~\cite{Kardos:2014zba}. Furthermore, NLO predictions have been provided in the narrow-width approximation~\cite{Melnikov:2011ta,Bevilacqua:2019quz,Stremmer:2024ecl} and with full off-shell effects~\cite{Bevilacqua:2018woc,Bevilacqua:2018dny,Stremmer:2024zhd}. More recently, approximate NNLO results have become available. One approach incorporates second-order soft-gluon corrections into the complete NLO result~\cite{Kidonakis:2022qvz}, while another~\cite{Wang:2025kpk} employs the massification procedure~\cite{Mitov:2006xs} together with a semi-massive scheme~\cite{Wang:2024pmv}, in which all loop integrals entering the amplitudes are treated as massless. The first steps towards an exact NNLO QCD calculation have also been taken with the computation of the one-loop amplitudes through $\mathcal{O}(\eps^2)$~\cite{Bera:2025upg}. The main bottleneck in reaching NNLO QCD accuracy for $pp \to t\bar{t}\gamma$ is the computation of the two-loop five-particle scattering amplitudes that enter the double-virtual corrections. 
This is particularly challenging because of the large number of kinematic variables and the complicated analytic structures arising from the multiple massive internal propagators in the loop integrals.

Thanks to recent advances in both integral reduction and the computation of loop integrals, two-loop $2\to3$ amplitudes with massive internal propagators have come within reach~\cite{Badger:2022hno,FebresCordero:2023pww,Badger:2024fgb,Agarwal:2024jyq,Becchetti:2025oyb,Badger:2025ilt,Badger:2025ljy,Becchetti:2025qlu,Gotzfried:2026def,Becchetti:2026awn,Becchetti:2026yxl}.
Only in the past year have the first NNLO QCD results making use of two-loop amplitudes with exact mass dependence appeared, for $t\bar{t}+$jet~\cite{Badger:2025ilt,Badger:2025ljy}, $t\bar{t}W$~\cite{Becchetti:2026awn,Becchetti:2026yxl} and the quark-initiated channel of $t\bar{t}H$~\cite{Chen:2026bxb} production.
These calculations have been performed in the generalised leading-colour approximation,
in which the two-loop finite remainders are expanded in the number of colours and of closed massless fermion loops, and only the leading terms are retained.
This approximation simplifies the calculation considerably, while still yielding a significant improvement in the precision of the resulting theoretical predictions.

In the case of $t\bar{t}\gamma$ production, the leading-colour two-loop amplitudes involve the three integral families already encountered in $t\bar{t}+$jet production, together with three additional families that contribute at leading colour because the photon carries no colour charge.
Here, by leading colour we mean the terms of the two-loop amplitudes that dominate when the number of colours, $N_c$, and the number of light-quark flavours, $n_l$, are large.
The amplitudes also contain terms of order $N_c \sum_f Q_f$, where the sum runs over the light-quark flavours and $Q_f$ is the fractional electric charge of flavour $f$.
Although in some cases these terms have been found to be comparable in size to the leading-colour ones (see e.g.\ ref.~\cite{Abreu:2023bdp}), they involve non-planar integrals that are much more challenging to compute, and we neglect them here.

The three new planar families are considerably more complicated than those relevant to $t\bar{t}+$jet production.
With up to three massive internal propagators, they exhibit a much richer analytic structure, in which several elliptic curves and nested square roots appear simultaneously.
To compute the integrals in these families, we follow the strategy developed in refs.~\cite{Badger:2024fgb,Badger:2024dxo} and successfully applied to $t\bar{t}+$jet and $t\bar{t}W$ production.
This strategy builds on the methods~\cite{Henn:2013pwa, Gehrmann:2018yef, Chicherin:2020oor, Chicherin:2021dyp, Abreu:2023rco} that underlie the rapid progress in two-loop $2\to3$ amplitudes with massless internal propagators over the past few years.
Specifically, we construct integral bases that satisfy algebraic differential equations of a form that allows us to express the solution in terms of a set of special functions, tailored to simplify the future computation of the two-loop leading-colour amplitudes.
We evaluate these special functions efficiently by numerically integrating the differential equations they satisfy~\cite{Boughezal:2006px, Czakon:2008zk, Czakon:2020vql, Badger:2025ljy}.

This article is organised as follows. In \cref{sec:Notation}, we introduce our notation and the relevant integral families. In \cref{sec:Bases_and_DEs}, we describe the structure of the differential equations and discuss how we obtained them, with emphasis on the sectors that involve elliptic curves or nested square roots. \Cref{sec:Solution} illustrates how we expressed the solution to the differential equations in terms of special functions. In \cref{sec:numerics}, we present our strategy for the evaluation of the special functions. 
We summarise our results and give an outlook in \cref{sec:Conclusions}. Appendix~\ref{app:New_Sectors} displays the five-point sectors of the three families studied here, and lists the corresponding master integrals in our bases.

\section{Notation}
\label{sec:Notation}

\begin{figure}[t]
\centering
\begin{subfigure}[t]{0.497\textwidth}
\centering
\includegraphics[scale=0.48]{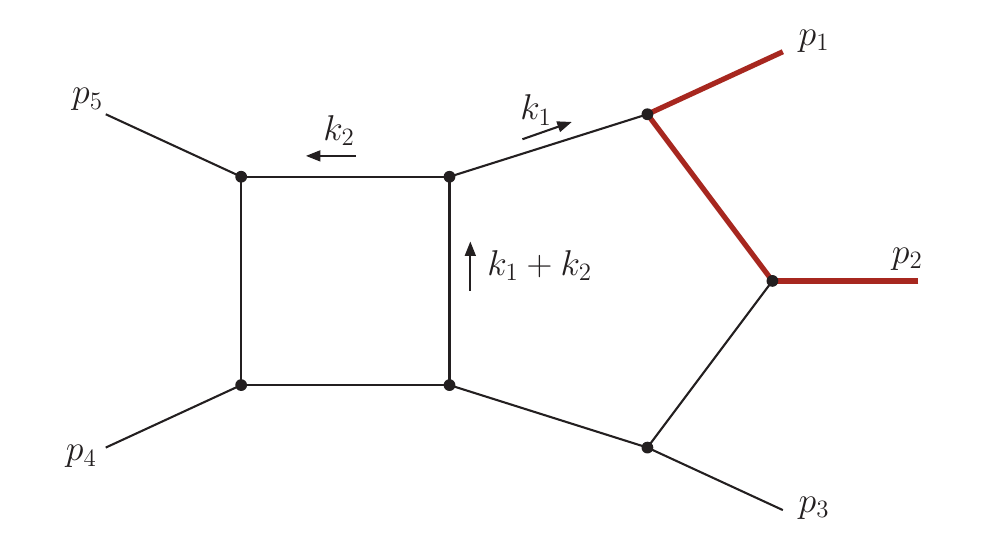}
\caption{\PBA, 88 MIs}
\end{subfigure}
\hfill
\begin{subfigure}[t]{0.496\textwidth}
\centering
\includegraphics[scale=0.48]{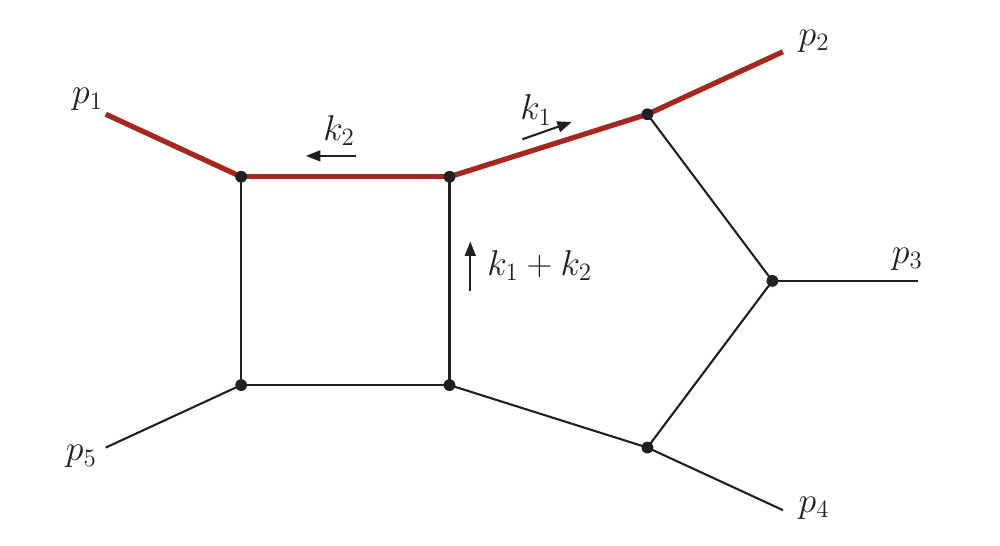}
\caption{\PBB, 121 MIs (1 ES, 1 NS)}
\end{subfigure}
\hfill
\begin{subfigure}[t]{0.496\textwidth}
\centering
\includegraphics[scale=0.48]{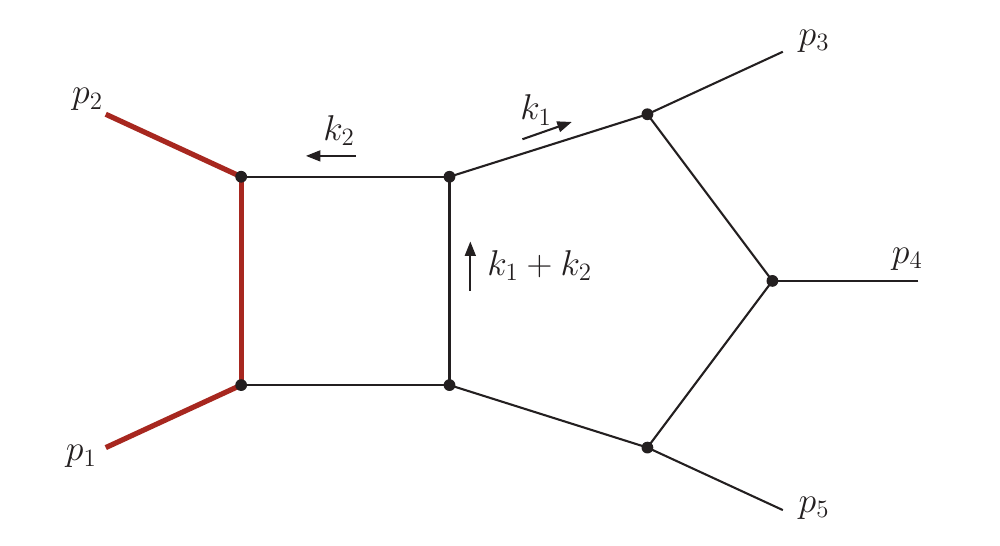}
\caption{\PBC, 109 MIs}
\end{subfigure}
\caption{Two-loop integral families contributing to both $t \bar{t} \gamma$ and $t \bar{t}+$jet production at hadron colliders in the leading-colour approximation, together with the corresponding number of master integrals and, in parentheses, their noteworthy analytic features. 
Red lines correspond to (anti-)top quarks, while black lines denote massless particles.
These families were computed in refs.~\cite{Badger:2022hno, Badger:2024fgb,Badger:2024dxo, Becchetti:2025oyb}.
We use the abbreviation ES to denote an elliptic sector, i.e.\ a sector associated with a single elliptic curve, and NS to denote a nested-square-root sector, i.e.\ a sector associated with a doublet of nested square roots.}
\label{fig:ttj_families}
\end{figure}

\begin{figure}[t]
\centering
\begin{subfigure}[t]{0.497\textwidth}
\centering
\includegraphics[scale=0.48]{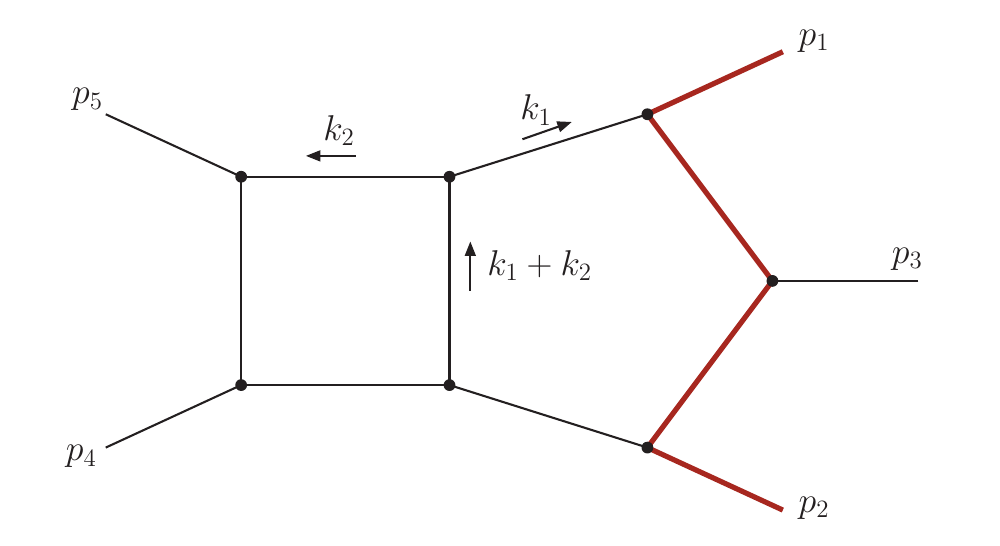}
\caption{\PBD, 100 MIs}
\end{subfigure}
\hfill
\begin{subfigure}[t]{0.496\textwidth}
\centering
\includegraphics[scale=0.48]{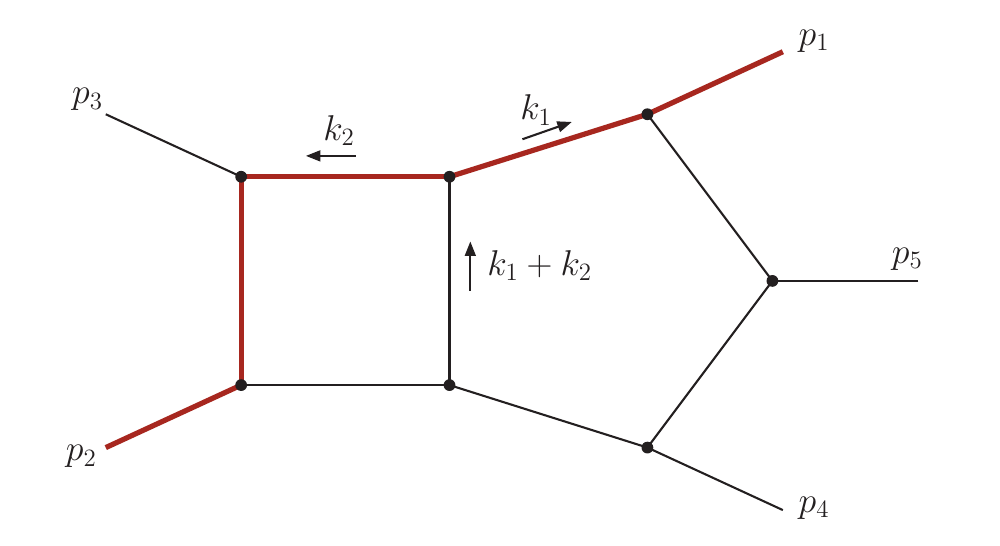}
\caption{\PBE, 154 MIs (2 ES, 1 NS)}
\end{subfigure}
\hfill
\begin{subfigure}[t]{0.496\textwidth}
\centering
\includegraphics[scale=0.48]{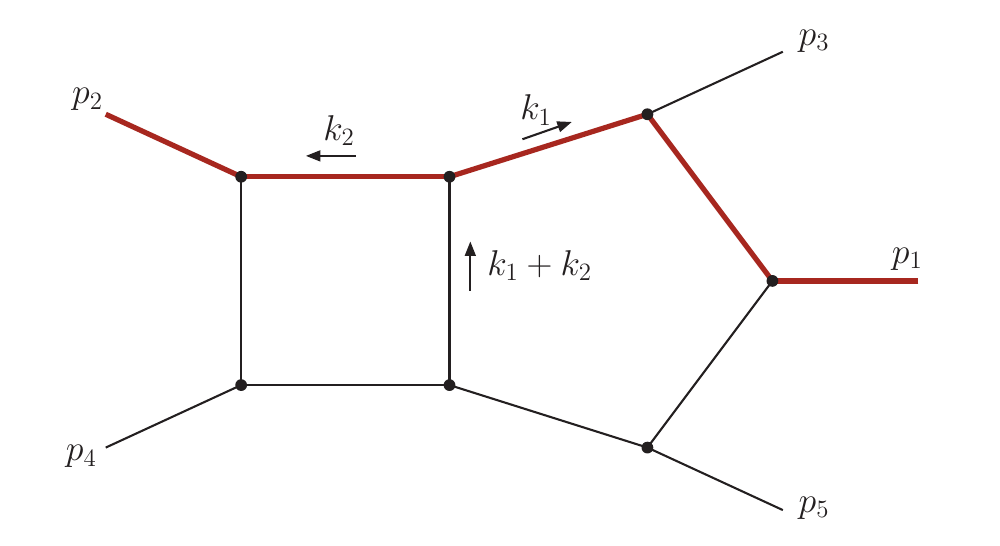}
\caption{\PBF, 142 MIs  (3 ES, 2 NS)}
\end{subfigure}
\caption{The two-loop integral families computed for the first time in this work.
The notation is the same as in \cref{fig:ttj_families}.}
\label{fig:tta_families}
\end{figure}

We generated the Feynman diagrams for the two-loop partonic amplitudes contributing to $t\bar{t}\gamma$ production at hadron colliders with \textsc{QGRAF}~\cite{Nogueira:1991ex}, and retained only the leading terms in the expansion in the number of colours and of light-quark flavours circulating in the loops. The Feynman integrals that appear can be grouped into six planar families.
Three of these families~---~\PBA, \PBB\ and \PBC, shown in \cref{fig:ttj_families}~---~are also needed for the production of a top-quark pair in association with a jet in the same approximation, and were computed in refs.~\cite{Badger:2022hno, Badger:2024fgb,Badger:2024dxo, Becchetti:2025oyb}. 
Here we focus on the remaining three families: \PBD, \PBE\ and \PBF,  depicted in \cref{fig:tta_families}. In addition to these six pentagon-box topologies, there are also hexagon-triangle topologies, which can be reduced either to the pentagon-boxes shown in \cref{fig:ttj_families,fig:tta_families} or to products of one-loop integrals.
We follow the conventions of ref.~\cite{Badger:2024fgb}, so that the results for all $t\bar{t}\gamma$ families can be combined straightforwardly.

We consider scattering processes of the form
\begin{equation} \label{eq:process}
P(-p_4)\,P'(-p_5) \to t(p_1)\,\bar{t}(p_2)\,\gamma(p_3)\, ,
\end{equation}
where the incoming partons $P$ and $P'$ can be either two gluons or a quark--antiquark pair. 
The external momenta $p_i$ are four-dimensional and all outgoing, so that momentum conservation reads
\begin{equation}
\label{mom_cons}
p_1 + p_2 + p_3 + p_4 + p_5 = 0 \, .
\end{equation}
The on-shell conditions are given by 
\begin{equation}
\label{onshell}
p_1^2 = p_2^2 = m_t^2 \,, \qquad \qquad p_3^2 = p_4^2 = p_5^2 = 0 \, ,
\end{equation}
where $m_t$ is the top-quark mass.
There are six independent Lorentz invariants, which we choose as
\begin{equation}
\label{invariants}
\vec{x} = \left(d_{12}, d_{23}, d_{34}, d_{45}, d_{15}, m_t^2 \right) \, ,
\end{equation}
with $d_{ij} = p_i \cdot p_j$. 
In the following, we will occasionally use the other scalar products to simplify the expressions.
For convenience, we write them here in terms of~$\vec{x}$:
\begin{equation}
\label{eq:other_invariants}
\begin{split}
d_{13} &= d_{45} - d_{12} - d_{23} - m_t^2 \,,\\
d_{14} &= d_{23} - d_{15} - d_{45} \,, \\
d_{24} &= d_{15} - d_{23} - d_{34} \,, \\
d_{25} &= d_{34} - d_{12} - d_{15} - m_t^2 \,, \\
d_{35} &= d_{12} - d_{34} - d_{45} + m_t^2 \,.
\end{split}
\end{equation}
The physical region of interest for $t\bar{t}\gamma$ production is the $s_{45}$ channel.
The conditions for the external momenta to be real and correspond to positive energies and physical scattering angles are given by the following inequalities in the scalar invariants,
\begin{equation}
\label{eq:physical_region}
\begin{aligned}
\quad d_{12}>0 \,, \qquad  d_{23}>0 \,, \qquad d_{34}<0\,, \qquad d_{45}>0 \,, \qquad d_{15}<0 \,, \\
d_{13}>0 \,, \qquad d_{14}<0 \,, \qquad d_{24}<0 \,, \qquad d_{25}<0 \,, \qquad d_{35}<0 \,,
\end{aligned}
\end{equation}
supplemented by $m_t^2 >0$ and the Gram-determinant constraints
\begin{equation}
\label{eq:physical_region2}
\mathbb{G}(p_i,p_j)<0 \,, \quad \mathbb{G}(p_i,p_j,p_k)>0 \,, \quad \mathbb{G}(p_i,p_j,p_k,p_l)<0 \,,
\end{equation}
where $i$, $j$, $k$ and $l$ take distinct values in $\{1,2,3,4,5\}$.
In \cref{eq:physical_region2}, $\mathbb{G} $ denotes the Gram determinant, defined for arbitrary four-vectors $a_i$ and $b_i$ as
\begin{equation}
\label{eq:gram-definitions}
\mathbb{G}(a_1,\ldots,a_n ) = \mathbb{G} \begin{pmatrix} a_1,\ldots,a_n \\ a_1,\ldots,a_n \end{pmatrix}\,, \qquad  \quad \mathbb{G}\begin{pmatrix} a_1,\ldots,a_n \\ b_1,\ldots,b_n \end{pmatrix}  = {\text{det}}\left(a_i\cdot b_j \right) \,.
\end{equation}

The integrals of each family $\mathrm{F} \in \{ \PBD, \PBE, \PBF \}$ take the form
\begin{equation}
\label{eq:family}
\begin{aligned}
I^{(\mathrm{F})}_{\vec{\nu}} \left[N \right] & = \int \frac{\dd^{d} k_1 \, e^{ \eps \gamma_{\rm E}}}{\ii \pi^{\frac{d}{2}}} \frac{\dd^{d} k_2 \, e^{ \eps \gamma_{\rm E}}}{\ii \pi^{\frac{d}{2}}}  \, N \, \prod_{i=1}^{11} \frac{1}{D^{\nu_i}_{\mathrm{F},i}} \,, \\
I^{(\mathrm{F})}_{\vec{\nu}}  & \equiv I^{(\mathrm{F})}_{\vec{\nu}} \left[1 \right] \,,
\end{aligned}
\end{equation}
where $\vec{\nu}=\left(\nu_1,\ldots, \nu_{11}\right) \in \mathbb{Z}^{11}$, $N\equiv N(k_i,p_j)$ is a polynomial in scalar products of external and loop momenta, $d = 4 - 2\eps$ is the space-time dimension, and the inverse propagators $D_{\mathrm{F},i}$ are defined in \cref{tab:propagators}.
The last three inverse propagators of each family are irreducible scalar products (ISPs), and thus their indices~---~$\nu_9$, $\nu_{10}$ and $\nu_{11}$~---~are always non-positive.
Hereafter, the family label $\mathrm{F}$ on the inverse propagators $D_{\mathrm{F},i}$ will be suppressed whenever it is clear from the context.
In \cref{eq:family}, we have omitted the dependence on $\vec{x}$ and $\eps$ for the sake of brevity.
\begin{table}[h!]
\centering
\begin{tabular}{c|c|c|c}
$\mathrm{F}$ & \PBD & \PBE & \PBF \\
\hline
$D_{\mathrm{F},1}$ & $k_1^2$ & $k_1^2-m_t^2$ & $k_1^2-m_t^2$ \\
$D_{\mathrm{F},2}$ & $(k_1-p_1)^2-m_t^2$ & $(k_1-p_1)^2$ & $(k_1-p_3)^2-m_t^2$ \\
$D_{\mathrm{F},3}$ & $(k_1-p_{13})^2-m_t^2$ & $(k_1+p_{234})^2$ & $(k_1-p_{13})^2$ \\
$D_{\mathrm{F},4}$ & $(k_1-p_{123})^2$ & $(k_1+p_{23})^2$ & $(k_1+p_{24})^2$ \\
$D_{\mathrm{F},5}$ & $k_2^2$ & $k_2^2-m_t^2$  & $k_2^2-m_t^2$ \\
$D_{\mathrm{F},6}$ & $(k_2+p_{123})^2$ & $(k_2-p_{3})^2-m_t^2$ & $(k_2-p_2)^2$ \\
$D_{\mathrm{F},7}$ & $(k_2+p_{1234})^2$ & $(k_2-p_{23})^2$ & $(k_2-p_{24})^2$ \\
$D_{\mathrm{F},8}$ & $(k_1+k_2)^2$ & $(k_1+k_2)^2$ & $(k_1+k_2)^2$ \\
$D_{\mathrm{F},9}$ & $(k_1-p_{1234})^2$ & $(k_1+p_{3})^2-m_t^2$ & $(k_1+p_2)^2$ \\
$D_{\mathrm{F},10}$ & $(k_2+p_1)^2-m_t^2$ & $(k_2+p_1)^2$ & $(k_2+p_3)^2-m_t^2$ \\
$D_{\mathrm{F},11}$ & $(k_2+p_{13})^2-m_t^2$ & $(k_2-p_{234})^2$ & $(k_2+p_{13})^2$ \\
\end{tabular}
\caption{Inverse propagators $D_{\mathrm{F},i}$ in \cref{eq:family} for the families computed in this work. 
We used the shorthand $p_{i\cdots j} = p_i + \cdots + p_j$.
}
\label{tab:propagators}
\end{table}

For each family, we determined a set of linearly independent master integrals (MIs) by solving systems of integration-by-parts (IBP)~\cite{Tkachov:1981wb,Chetyrkin:1981qh,Laporta:2001dd} and symmetry relations.
We used \textsc{LiteRed}~\cite{Lee:2012cn, Lee:2013mka} and \textsc{NeatIBP}~\cite{Wu:2023upw} to generate the IBP systems, and \textsc{FiniteFlow}~\cite{Peraro:2019svx} to solve them with finite-field methods~\cite{vonManteuffel:2014ixa,Peraro:2016wsq}.
We found 100, 154 and 142 MIs for \PBD, \PBE\ and \PBF, respectively. 
After excluding MIs already available in the literature~\cite{Badger:2022hno, Badger:2024fgb, Becchetti:2025oyb,FebresCordero:2023pww,Becchetti:2025qlu}, and modding out by permutations of the external momenta as well as products of one-loop integrals, we obtain 46 new, genuinely two-loop five-point master integrals distributed across 14 sectors. These have not been computed previously and constitute the new result of our work. We list these sectors and their MIs in appendix~\ref{app:New_Sectors}.
The families considered here also contain several four-point subsectors that have not been computed before. As these are simpler, we do not discuss them explicitly.

\section{Integral bases and differential equations}
\label{sec:Bases_and_DEs}

In order to compute the MIs of each family, we view them as the solutions to systems of differential equations (DEs) in the kinematic invariants~\cite{Barucchi:1973zm, Kotikov:1990kg, Kotikov:1991hm, Bern:1993kr, Gehrmann:1999as}.
In this section, we discuss the construction of bases of MIs that satisfy DEs well suited for an efficient solution.
The numerical evaluation of the solution is deferred to \cref{sec:numerics}.

The derivatives of an arbitrary set of MIs, $\vec{\mathcal{J}}^{(\mathrm{F})} (\Vec{x},\eps)$, are given by linear combinations of scalar integrals of the same family $\mathrm{F}$.
IBP-reducing these to the MIs themselves thus yields a system of first-order partial DEs for the MIs,
\begin{equation}
\label{eq:DEs_general}
\frac{\partial}{\partial \xi} \vec{\mathcal{J}}^{(\mathrm{F})} (\Vec{x},\eps) = \mathcal{A}_\xi^{({\mathrm{F}})}(\Vec{x},\eps)\cdot \vec{\mathcal{J}}^{(\mathrm{F})} (\Vec{x},\eps),  \qquad \forall \ \xi \in \Vec{x}\,.
\end{equation}
Equivalently, all partial DEs may be combined in a single line, as
\begin{equation}
\label{eq:DEs_general_2}
\dd \vec{\mathcal{J}}^{(\mathrm{F})} (\Vec{x},\eps) =\dd \mathcal{A}^{(\mathrm{F})}(\Vec{x},\eps) \cdot \vec{\mathcal{J}}^{(\mathrm{F})} (\Vec{x},\eps) \,,
\end{equation}
where $\dd$ is the total differential in the kinematic invariants $\vec{x}$, and
\begin{equation}
\dd \mathcal{A}^{({\mathrm{F}})}(\Vec{x},\eps)=\sum_{\xi \in {\vec{x}}} \mathcal{A}_\xi^{(\mathrm{F})}(\Vec{x},\eps) \, \dd\xi \, .
\end{equation}
For an arbitrary basis $\vec{\mathcal{J}}^{(\mathrm{F})}$, the connection matrix $\dd \mathcal{A}^{(\mathrm{F})}(\Vec{x},\eps)$ in the DEs~\eqref{eq:DEs_general_2} may be complicated, rendering their solution or even derivation challenging.
This problem can be greatly simplified by a suitable choice of basis.
A particularly successful approach is to obtain the \emph{canonical form}~\cite{Henn:2013pwa},
\begin{equation}
\label{eq:eps-fact}
\dd \vec{\mathcal{J}}^{(\mathrm{F})} (\Vec{x},\eps) = \eps \, \dd \tilde{\mathcal{A}}^{(\mathrm{F})}(\Vec{x}) \cdot \vec{\mathcal{J}}^{(\mathrm{F})} (\Vec{x},\eps) \,,
\end{equation}
where $\eps$ factors out of the connection matrix, and the latter has only simple poles
near all singular loci.
In the best-understood case, $\dd \tilde{\mathcal{A}}^{(\mathrm{F})}(\Vec{x})$ is further simplified to a linear combination of logarithmic one-forms called $\dd \log$ forms,
\begin{align}
\label{eq:dlog}
  \dd \tilde{\mathcal{A}}^{(\mathrm{F})}(\Vec{x}) = \sum_i \alpha_i^{(\mathrm{F})} \, \dd \log W_i(\vec{x}) \,,
\end{align}
where the $\alpha_i^{(\mathrm{F})}$'s are constant rational matrices, and the $W_i$'s are algebraic functions called \emph{letters}.
The DE form in \cref{eq:eps-fact,eq:dlog} implies that the corresponding MIs are \emph{pure} transcendental functions, and enables the use of powerful mathematical techniques to handle and evaluate the solution efficiently.

Among the families considered in this work, achieving the form in \cref{eq:eps-fact,eq:dlog} is possible only for \PBD.
The remaining families exhibit a considerably more complex analytic structure.
In particular, the presence of elliptic curves implies that, although the $\eps$-factorised form in \cref{eq:eps-fact} may exist, achieving it is substantially more difficult and the resulting connection matrices would contain transcendental functions rather than algebraic logarithmic derivatives as in \cref{eq:dlog}.
In addition, some leading singularities~---~the maximal multivariate residues of the integrand~\cite{Cachazo:2008vp,Arkani-Hamed:2010pyv}~---~contain doublets of nested square roots,
\begin{align} \label{eq:NSform}
n_{\pm}(\vec{x}) = \sqrt{A(\vec{x}) \pm B(\vec{x}) \, \sqrt{C(\vec{x})}} \,,
\end{align}
where $A$, $B$ and $C$ are rational functions of $\vec{x}$.
It is not known whether nested square roots can generally be accommodated into logarithmic one-forms as in \cref{eq:dlog} and, even when they can, doing so requires new algorithms~\cite{FebresCordero:2023pww,Becchetti:2025oyb,Becchetti:2025qlu,Gotzfried:2026def}.
While canonical DEs for \PBB\ have recently been obtained~\cite{Becchetti:2025oyb}~---~the first time for a family of five-point integrals containing an elliptic curve~---~evaluating the solution with the efficiency required for phenomenological applications in the complex phase space of a $2\to 3$ process remains challenging.
Moreover, the families considered here are significantly more intricate than \PBB: \PBE\ involves two elliptic curves and a doublet of nested square roots, while \PBF\ contains three elliptic curves and two such doublets.
The characteristic analytic features of each family are summarised in \cref{fig:ttj_families,fig:tta_families}.

For these reasons, we adopt the strategy of refs.~\cite{Badger:2024fgb,Becchetti:2025qlu} for the construction of the integral bases.
The guiding principle is to retain algebraic DEs while relaxing the requirement of full $\eps$ factorisation, and to keep the deviation from the canonical form as small as possible.
The structure of the resulting DEs is discussed in \cref{sec:DEs}.
We further devote \cref{sec:polylogarithmic_secs,sec:elliptics,sec:nesteds} to the construction of integral bases for the logarithmic, nested-square-root and elliptic sectors, respectively.
We indicate each sector by its corner integral.

\subsection{Form of the differential equations}
\label{sec:DEs}

We obtain DEs with connection matrices of the following form,
\begin{equation}
\label{eq:our_connection_matrices}
\dd \mathcal{A}^{(\mathrm{F})}(\Vec{x},\eps) = \sum_{i=0}^2 \eps^i \left( \sum_{j} \alpha_{j}^{(\mathrm{F},i)}  \dd \log \bigl(W_j(\vec{x})\bigr) + \sum_j \beta_{j}^{(\mathrm{F},i)} \omega_j(\vec{x}) \right) \, ,
\end{equation}
where $\alpha_{j}^{(\mathrm{F},i)}$ and $\beta_{j}^{(\mathrm{F},i)}$ are constant rational matrices.
Compared with the canonical form in \cref{eq:eps-fact,eq:dlog}, the dependence on $\eps$ is quadratic, and non-$\dd\log$ one-forms, named $\omega_j(\vec{x})$, appear.

The non-zero entries in $\alpha_{j}^{(\mathrm{F},0)}$, $\alpha_{j}^{(\mathrm{F},2)}$ and $\beta_{j}^{(\mathrm{F},i)}$ capture the deviation from the canonical form.
The DEs are in canonical form for \PBD, i.e.\ $\alpha_{j}^{(\PBD,0)}=\alpha_{j}^{(\PBD,2)}=\beta_{j}^{(\PBD,i)}=0$.
For the remaining families, these matrices are sparse, and their non-zero entries roughly correspond to MIs associated with either elliptic curves or nested square roots.
To be more specific, we denote by elliptic (nested-square-root) sectors those whose maximal cuts (see the next sections) contain an elliptic curve (a doublet of nested square roots), and by logarithmic sectors those whose maximal cuts are free of such analytic features.
Then, the entries that couple MIs of logarithmic sectors are $\eps$-factorised and contain mostly $\dd\log$ one-forms.
If we further restrict to MIs of logarithmic sectors that do not couple to any elliptic or nested-square-root sectors, the corresponding entries are in canonical $\dd\log$ form.
The quadratic dependence on $\eps$ appears only in a subset of the entries associated with MIs of the elliptic sectors.
For the MIs of the nested-square-root sectors, the connection matrices are $\eps$-factorised in the entries coupling to logarithmic sectors, and linear in the entries coupling to elliptic sectors or to the nested-square-root sectors themselves.
We discuss the constraints satisfied by $\alpha_{j}^{(\mathrm{F},0)}$, $\alpha_{j}^{(\mathrm{F},2)}$ and $\beta_{j}^{(\mathrm{F},i)}$ for \PBE\ and \PBF\ in more detail in \cref{sec:Solution}, where we exploit them to solve the DEs order by order in $\eps$.

The DEs derived in this work contain the following square roots:
\begin{align}
\label{eq:sqroots}
\begin{aligned}
& r_1 = \sqrt{\mathbb{G}(p_1,p_2,p_3,p_4)} \,, \\
& r_2 = \sqrt{Y^{0|m_t|m_t|0}_{p_1|p_3|p_2|p_{45}}} \,, \\
& r_3 = \sqrt{d_{23}[d_{23} (d_{25}+d_{45})^2 -2d_{45} d_{35} m_t^2]} \,, \\
& r_4 = \sqrt{d_{13} [d_{13} (d_{14} + d_{45})^2 - 2 d_{45} d_{34} m_t^2]} \,, \\
& r_5 = \sqrt{-\mathbb{G}(p_{13},p_2)} \,, \\
& r_6 = \sqrt{-\mathbb{G}(p_{45},p_1)} \,, \\
& r_7 = \sqrt{d_{45}(2d_{23}-m_t^2)[d_{45}(2d_{23}+3m_t^2)-2(d_{23}+m_t^2)f_{123}]+d_{23}^2 f_{123}^2} \,, \\
& r_8 = \sqrt{-\mathbb{G}(p_1,p_2)} \,, \\
& r_9 = \sqrt{-\mathbb{G}(p_{15},p_2)} \,, \\
& r_{10} = \sqrt{f_{234}^2 m_t^4+d_{15}^2 (m_t^2-2d_{23})^2+2d_{15}m_t^2[2d_{23}f_{234}+(d_{34}-d_{23})m_t^2]} \,, \\
& r_{11} = \sqrt{-\mathbb{G}(p_{24},p_1)}\,, \\
& r_{12} = \sqrt{f_{135}^2 m_t^4+d_{24}^2 (m_t^2-2d_{13})^2+2d_{24} m_t^2[2d_{13}f_{135}+(d_{35}-d_{13})m_t^2]} \,, \\
& r_{13} = \sqrt{Y^{m_t|0|m_t|m_t}_{p_{13}|p_2|p_4|p_5}}\,,
\end{aligned}
\end{align}
where we introduced the determinant of the modified Cayley matrix,
\begin{equation}
\label{eq:Cayley}
Y^{m_1|m_2|m_3|m_4}_{P_1|P_2|P_3|P_4} = \det \left( \frac{1}{2} \left[(q_i - q_j)^2 - m_i^2 - m_j^2 \right] \right) \,, \qquad q_i = \sum_{k=0}^{i-1} P_k \,,
\end{equation}
for $i,j = 1, \ldots , 4$, with $P_0 = 0$, as well as the shorthands $p_{ij}=p_i+p_j$ and
\begin{align}
f_{123}=2(d_{12}+d_{23})+m_t^2 \,, \qquad f_{234}=d_{23}+d_{34}\,, \qquad f_{135}=d_{13}+d_{35} \,.
\end{align}
We further recall that the Gram determinants are defined in \cref{eq:gram-definitions}. 
We take the principal branch of all square roots.

Since the IBP relations have rational coefficients, square roots enter the otherwise rational DEs only through the definition of the MIs, where they appear as overall normalisation factors (see \cref{sec:nesteds,sec:elliptics} for examples).
As a result, every MI and every one-form, whether $\dd\log$ or not, is either even or odd under flipping the sign of each square root.
We say that an object has \emph{charge} $r$ if it is odd under $r \to -r$.
For instance, if two MIs have charges $r_i$ and $r_j$ respectively, with $i\neq j$, the entry of the connection matrix coupling the derivative of one to the other has charge $r_i r_j$.
This structure is useful for two reasons: it constrains the form of the one-forms, and it splits them into subsets of different charge which cannot be related to one another by $\mathbb{Q}$-linear relations, and can therefore be analysed separately.
Nested square roots of the form in \cref{eq:NSform} complicate this picture, as they introduce doublets of objects that transform into each other~\cite{Becchetti:2025oyb}.
As discussed above, we prefer to avoid the introduction of nested square roots even though they appear in the leading singularities of some integrals (see \cref{sec:nesteds}).

The combined alphabet $\{W_i(\vec{x})\}$ of the \PBD, \PBE\ and \PBF\ families consists of 167 letters.
Of these, 67 are polynomials and have maximal total degree five.
The 100 algebraic letters are split into 39 charges, each involving a product of at most two of the square roots in \cref{eq:sqroots}.
The algebraic letters with charge $R$ take the form
\begin{equation}
\label{eq:algebraic_letters}
W(\vec{x})=\frac{q(\vec{x})+Q(\vec{x}) R(\vec{x})}{q(\vec{x})-Q(\vec{x}) R(\vec{x})} \,,
\end{equation}
where $q(\vec{x})$ and $Q(\vec{x})$ are polynomials in $\vec{x}$.
The form in \cref{eq:algebraic_letters} makes it manifest that $\dd\log W$ is odd with respect to $R$.
To determine the alphabet, we employed \textsc{BaikovLetters}~\cite{Jiang:2024eaj} and \textsc{SOFIA}~\cite{Caron-Huot:2024brh, Correia:2025wtb}, supplemented by a direct analysis of the denominators appearing in the DEs.
This procedure allowed us to identify the complete rational alphabet and the majority of the algebraic letters.
We obtained the missing algebraic letters either by applying the algorithm of refs.~\cite{Heller:2019gkq,Zoia:2021zmb,Matijasic:2024too} or by explicitly integrating selected entries of the DEs and rewriting the resulting expressions in the form of \cref{eq:algebraic_letters}.
Two rational letters associated with the elliptic sectors $I^{(\PBE)}_{10101111000}$ and $I_{11101011000}^{(\PBF)}$, namely
\begin{equation}
\begin{split}
W_{38}(\vec{x}) &= d_{15} d_{23} (d_{23} + d_{34} - 2 m_t^2) + 2 d_{15}^2 m_t^2 + d_{23} (d_{23} + d_{34}) m_t^2 \,, \\
W_{57}(\vec{x}) &= \left. W_{38}(\vec{x}) \right|_{p_1 \leftrightarrow p_2}^{p_4 \leftrightarrow p_5} \,,
\end{split} 
\end{equation}
were not predicted by either \textsc{BaikovLetters} or \textsc{SOFIA}.

Non-$\dd\log$ one-forms are instead one-forms that are not written as the differential of a logarithm.
Across \PBE\ and \PBF, there are 389 $\mathbb{Q}$-linearly independent non-$\dd\log$ one-forms.
The basis $\{\omega_i(\vec{x})\}$ in \cref{eq:our_connection_matrices} is chosen from the entries of the connection matrices by preferring those with lower total polynomial degree in the numerator.
All but four non-$\dd\log$ one-forms are not closed, which implies that they cannot be written as the differential of any function.
Of the four closed non-$\dd\log$ one-forms, one has a rational primitive and is related to the elliptic sector $I^{(\PBF)}_{11011101000}$ (discussed in \cref{sec:elliptics}).
The remaining three closed one-forms originate from off-diagonal entries in the nested-square-root sectors described in \cref{sec:nesteds}; they can be written as $\dd\log$ forms, but the arguments would contain nested square roots, which we prefer to avoid.
Instead, we adopt a local representation in terms of differentials of the variables, as
\begin{align} \label{eq:omega}
 \omega_i(\vec{x}) = R_i(\vec{x}) \, \sum_{j=1}^6 \omega_{ij}(\vec{x}) \, \dd x_j \,,
\end{align}
where $R_i(\vec{x})$ is the relevant charge and the $\omega_{ij}(\vec{x})$'s are rational.
The non-$\dd\log$ one-forms feature 24 distinct charges (including $R_i = 1$), each involving at most two square roots. 
One of these charges does not figure among those of the algebraic letters. 
To simplify the expressions, we decomposed the coefficients $\omega_{ij}(\vec{x})$ into partial fractions with \textsc{MultivariateApart}~\cite{Heller:2021qkz}.
All irreducible polynomials appearing in the denominators either belong to the alphabet, are predicted by \textsc{SOFIA}, or arise from the discriminants of the elliptic curves.
The highest-degree denominator polynomial has total degree five.
The numerators of the partial fractions are generally more involved, with the highest-degree polynomial reaching degree 25.
We wrote all the partial fractioned one-forms in terms of multiplicatively independent polynomials to reduce the number of arithmetic operations (see also section~\ref{sec:numerics}) and used \textsc{Mathematica}'s \texttt{Simplify} function to simplify the independent factors.
These expressions are substantially more complicated than those for \PBA, \PBB\ and \PBC~\cite{Badger:2024fgb}, reflecting the richer analytic structure. 
For instance, when the one-forms are written over a common denominator, the maximal total degree of the polynomials they contain is 25, compared with 14 in ref.~\cite{Badger:2024fgb}.

\subsection{Logarithmic sectors}
\label{sec:polylogarithmic_secs}

Identifying MIs that obey canonical DEs for the logarithmic sectors of a family is, in general, a non-systematic task often requiring a combination of different techniques and insights. 
In our construction, we first determined the subsectors in common with families already studied in the literature~\cite{Badger:2022hno, Badger:2024fgb, FebresCordero:2023pww}, and adopted the corresponding pure MIs. 
In order to obtain a suitable basis for \PBD, we took the basis provided in ref.~\cite{FebresCordero:2023pww} for their $T_1$ family, set the Higgs-boson mass to zero at the integrand level, removed the integrals that become redundant in the limit, and made a few manual adjustments. 

For the genuinely new MIs of \PBE\ and \PBF, we proceeded bottom-up, sector by sector. 
For each sector, we initially considered the maximal cut, neglecting the connection-matrix entries coupling to lower sectors, and searched for a \emph{pre-canonical basis} of integrals that are either pure or render the DEs linear in $\eps$. 
We rotated the latter into pure integrals by the Magnus exponential method~\cite{Argeri:2014qva, Gehrmann:2014bfa}. 
Having obtained a basis that is pure on the maximal cut, we released the cut and studied the subsectors. 
In most cases, the connection-matrix entries coupling to subsectors were linear in $\eps$, and the $\eps^0$ terms could be straightforwardly integrated out with the Magnus exponential method. 
In some cases, however, the dependence on $\eps$ in the subsector entries was more involved, and we replaced some of the MIs that were pure on the maximal cut with alternative choices until linearity in $\eps$ was obtained also for the subsectors.

We determined the pre-canonical bases through a combination of two complementary approaches. 
The first consists of an integrand analysis~\cite{Cachazo:2008vp, Arkani-Hamed:2010pyv, Henn:2020lye, Dlapa:2021qsl} in the Baikov representation~\cite{Baikov:1996rk, Baikov:1996iu, Frellesvig:2017aai}, aimed at identifying MIs with only simple poles and constant leading singularities. 
For this purpose, we employed the package \textsc{Baikov.m}~\cite{Frellesvig:2017aai} to generate the Baikov representation and \textsc{DLogBasis}~\cite{Henn:2020lye} to analyse it. 
The second approach relies on educated guesses based on analogous sectors appearing in previously studied integral families, followed by explicit checks that the corresponding DEs are linear in $\eps$ on the maximal cut. 
These checks are computationally inexpensive, as they require reconstructing only the dependence on $\eps$, while keeping the kinematic variables $\vec{x}$ fixed to random values in a finite field. 
For sectors with five or fewer propagators and at most four-point kinematics, integrals with dotted propagators~---~i.e.\ with propagator power $2$~---~proved to be suitable choices. 
For sectors with four-point kinematics and six or more propagators, we generally preferred MIs with numerators. 
Finally, for the five-point sectors, we also used numerator insertions involving the $\eps$-dimensional components of the loop momenta.
These are conventionally denoted as $\mu_{ij} = k^{[-2\eps]}_i  \cdot k^{[-2\eps]}_j$, where $k_i=k^{[4]}_i+k^{[-2\eps]}_i$ with $k^{[-2\eps]}_i \cdot k^{[4]}_j = 0 = k^{[-2\eps]}_i \cdot p_j$, and can be expressed in terms of propagators of a given family via Gram determinants, as
\begin{align}
 \mu_{ij} =  \frac{\mathbb{G} \begin{pmatrix} k_i,p_1,p_2,p_3,p_4 \\ k_j,p_1,p_2,p_3,p_4 \end{pmatrix}}{\mathbb{G}(p_1,p_2,p_3,p_4)}  \,.
\end{align}

\subsection{Nested square roots}
\label{sec:nesteds}

\begin{figure}[t]
\centering
\begin{subfigure}[b]{0.496\textwidth}
\centering
\includegraphics[scale=0.58]{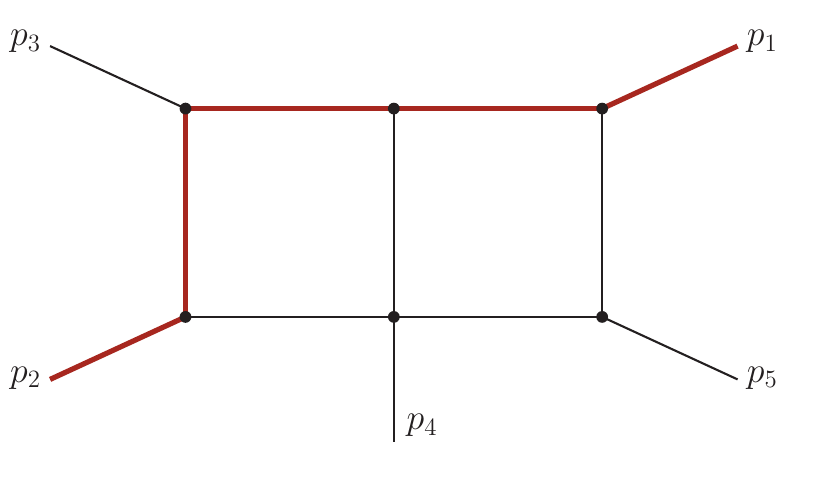}
\caption{Sector $I_{11101111000}^{(\PBE)}$ (MIs \#5--11)}
\end{subfigure}
\hfill
\begin{subfigure}[b]{0.496\textwidth}
\centering
\includegraphics[scale=0.52]{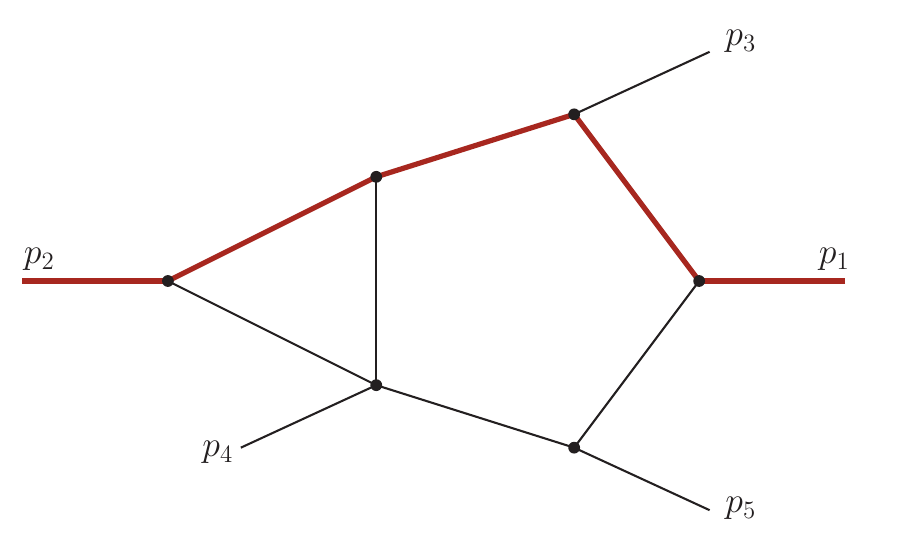}
\caption{Sector $I_{11111101000}^{(\PBF)}$ (MIs \#17--19)}
\end{subfigure}
\caption{The two independent nested-square-root sectors appearing in the integral families under study and, in parentheses, the position of the corresponding MIs in our bases.}
\label{fig:nested-sectors}
\end{figure}

Three nested-square-root sectors appear in the families considered in this work:
\begin{align*}
 I_{11101111000}^{(\PBE)} \,, \quad I_{11111101000}^{(\PBF)}\,, \quad I_{11101111000}^{(\PBF)} \,.
\end{align*}
All three sectors have five-point kinematics, and $I_{11101111000}^{(\PBF)}$ coincides with $I_{11101111000}^{(\PBE)}$ upon exchanging $p_1 \leftrightarrow p_2$ and $p_4 \leftrightarrow p_5$.
We display two independent sectors in \cref{fig:nested-sectors}.

The sector $I_{11111101000}^{(\PBF)}$ has a pentagon-triangle topology and three MIs.
Other pentagon-triangle topologies with massive loop propagators have already been observed to give rise to nested square roots~\cite{Badger:2024fgb, Becchetti:2025qlu}.
The loop-by-loop Baikov integrand~\cite{Frellesvig:2017aai} of the corner integral is similar to the one encountered in ref.~\cite{Becchetti:2025qlu}, and we thus choose the MIs similarly:
\begin{equation}
\label{nested-PBF-basis}
\begin{split}
\mathcal{J}^{(\PBF)}_{17} &= \frac{\eps^4}{d_{45}} r_1 I_{11111101000}^{(\PBF)} \left[ 2 d_{15} d_{13} - d_{15} D_{1}- (d_{25}+d_{45}) D_{2} \right] \,, \\
\mathcal{J}^{(\PBF)}_{18} &=\frac{\eps^4}{d_{45}} I_{11111101000}^{(\PBF)} \left[ a_1 D_{9} -a_2 - a_3 D_{2}- a_4 D_{1} \right] \,, \\
\mathcal{J}^{(\PBF)}_{19} &= \eps^4 r_1 I_{11111101000}^{(\PBF)}[\mu_{11}] \,,
\end{split}
\end{equation}
where the $a_i$'s are polynomials in $\vec{x}$,
\begin{equation}
\begin{split}
a_1 &= d_{13} \left[(d_{12}-d_{35}) m_t^2-d_{12} (2d_{15}+m_t^2)+2d_{15}(d_{12}+d_{13})\right] \,, \\
a_2 &=2(d_{23}+d_{34})d_{45} (d_{13}d_{15}-d_{35} m_t^2)-2d_{13} d_{15} \left[ d_{13} d_{15}+d_{23} (d_{15}+d_{35})\right] \,, \\
a_3 &= d_{25}\left[d_{35}(d_{23}-d_{45})+d_{12}(d_{45}-d_{15})-d_{15} m_t^2\right]+d_{45}\left[d_{12} d_{45}-d_{15}(d_{24}+m_t^2)\right] \,, \\
a_4 &= d_{15}^2 (d_{23}+d_{13})-d_{35} d_{45} m_t^2+d_{15} \left[(d_{12}+m_t^2) (d_{23}-2d_{45})+(d_{45}-d_{23})(d_{34}+2d_{45})\right] \,.
\end{split}
\end{equation}
These polynomials appear in the Baikov representation of the corner integral, similarly to \cref{eq:nested_PBE_Baikov,eq:P1P2} for the other nested-square-root sector.
The integrals $\mathcal{J}^{(\PBF)}_{17}$ and $\mathcal{J}^{(\PBF)}_{18}$ are the ones associated with the nested square roots.
In particular, the basis defined in \cref{nested-PBF-basis} can be rotated into a form that is $\eps$-factorised in all entries except those coupling to the elliptic sectors, which remain linear in $\eps$, as follows
\begin{equation}
\label{rotation_nested_PBF}
\begin{pmatrix} \mathcal{J}^{(\PBF)}_{17} \\ \mathcal{J}^{(\PBF)}_{18} \end{pmatrix} \mapsto \frac{d_{45}}{a_1} \begin{pmatrix} n_1^+ & 0 \\ 0 & n_1^- \end{pmatrix} \cdot
\begin{pmatrix} 1 & 1 \\ -1 & 1 \end{pmatrix} \cdot \begin{pmatrix} \mathcal{J}^{(\PBF)}_{17} \\ \mathcal{J}^{(\PBF)}_{18} \end{pmatrix} \, ,
\end{equation}
where the doublet of nested square roots $n_1^{\pm}$ is given by
\begin{equation}
\label{nested_root_PBF}
n_1^{\pm}=\frac{\sqrt{(a_2 \pm  2 d_{15} d_{13} r_1)(a_2-4 a_1 m_t^2 \pm 2 d_{15} d_{13} r_1)}}{\pm r_1} \,.
\end{equation}
Nonetheless, we prefer to keep the basis in \cref{nested-PBF-basis} so that only simple square roots appear in the connection matrix, following refs.~\cite{Badger:2024fgb, Becchetti:2025qlu}.

We now turn to the sector $I_{11101111000}^{(\PBE)}$, which corresponds to a five-point double-box topology with three massive propagators and has seven MIs.
To the best of our knowledge, this is the first example of a topology of this type involving nested square roots. 
This, together with the relatively large number of MIs, makes the construction of a suitable basis more challenging for this sector. 
We analyse the corner integral in the loop-by-loop Baikov representation starting from the $k_2$-loop.
Cutting all propagators except for $D_{\PBE,1}$~---~an operation that we denote by $\text{M.C.}\setminus\{D_1\}$, where $\text{M.C.}$ stands for maximal cut~---~and setting $\eps=0$ yield
 \begin{equation}
\label{eq:nested_PBE_Baikov}
\left. I_{11101111000}^{(\PBE)} \right|^{\text{M.C.}\setminus\{D_1\}}_{\eps \to 0} \propto  \frac{a_1' \, r_1}{\sqrt{d_{23}}} \,  \int \dd z_1 \, \dd z_4 \, \dd z_9 \, \frac{1}{z_1 \, \sqrt{\mathcal{P}_1} \, \mathcal{P}_2 \, \overline{\mathcal{P}}_2} \, ,
\end{equation}
where we have omitted the overall kinematics-independent factors. 
In \cref{eq:nested_PBE_Baikov}, $\mathcal{P}_1$, $\mathcal{P}_2$, and $\overline{\mathcal{P}}_2$ are polynomials in the Baikov variables $z_1$, $z_4$ and $z_9$ (with $z_i = D_{\PBE,i}$), given~by
\begin{equation} \label{eq:P1P2}
\begin{split}
\mathcal{P}_1 & = d_{23} z_9^2 + 2z_4 (z_9-z_1) m_t^2 \,, \\
\mathcal{P}_2 & = a_1' z_9+ a_2' z_4+2 a_3' d_{45} + a_4' z_1+\left[d_{15}(z_4-2d_{45})+d_{45}z_1\right]r_1 \,, \\
\overline{\mathcal{P}}_2 & = \mathcal{P}_2 \bigr|_{r_1 \to -r_1}\,.
\end{split}
\end{equation}
In the equations above, the $a_i'$'s are polynomials in $\vec{x}$, defined by
\begin{equation}
\begin{split}
a_1' & = d_{45}\left[ 2d_{15}(d_{15}-d_{23}+d_{45})+d_{45} m_t^2 \right] \,,  \\
a_2' & = d_{35}(d_{15}^2+d_{45}m_t^2)+d_{15}\left[d_{15}d_{34}+(d_{23}-d_{45}) (d_{34}+2d_{45})-(d_{23}-2d_{45})(d_{12}+m_t^2)\right] \,, \\
a_3' & = d_{45} \left[d_{15} (2d_{23}+d_{34})+(d_{23}+d_{34})m_t^2\right]-d_{15}\left[d_{13}d_{15}+d_{23}(d_{12}-d_{15}+2d_{23}+d_{34}+m_t^2)\right]\,, \\
a_4' & = d_{23}d_{45}(d_{15}-d_{25})-d_{45}\left[d_{12}d_{15}+(d_{15}-d_{34}) (2d_{15}+d_{45})+(d_{15}+d_{45}) m_t^2\right] \,.
\end{split}
\end{equation}
In this analysis, we did not take the maximal cut and instead kept $D_{\PBE,1}$ in order to capture also the contributions from the subsectors in which $D_{\PBE,1}$ is an ISP.
These terms were necessary to render certain subsector entries of the connection matrix $\eps$-factorised.
Moreover, we noticed that leaving any other propagator uncut would introduce square roots involving the propagators appearing in $\mathcal{P}_2$ and $\overline{\mathcal{P}}_2$, yielding an integrand which cannot be trivially expressed in terms of loop integrals.
Based on the analysis of the integrand in \cref{eq:nested_PBE_Baikov}, we choose the following MIs:
\begin{equation}
\label{eq:nested-basis-PBE}
\begin{split}
%\mathcal{J}^{(\PBE)}_{5} &= \frac{\eps^4}{d_{45}} r_1 I_{11101111000}^{(\PBE)} \left[d_{45}  D_{\PBE,1} + d_{15} D_{\PBE,4} -2 d_{15} d_{45} \right] \,, \\
\mathcal{J}^{(\PBE)}_{5} &= \frac{\eps^4}{d_{45}} I_{11101111000}^{(\PBE)} \left[ \frac{\mathcal{P}_2 - \overline{\mathcal{P}}_2}{2} \right] \,, \\
%\mathcal{J}_{\PBE,6} &= \frac{\eps^4}{d_{45}} \left[ a_1' I_{11101111-100}^{(\PBE)}+ a_2'  I_{111-11111000}^{(\PBE)}+2 a_3' d_{45} I_{11101111000}^{(\PBE)}+a_4' I_{01101111000}^{(\PBE)}\right] \,, \\
\mathcal{J}^{(\PBE)}_{6} &= \frac{\eps^4}{d_{45}} I_{11101111000}^{(\PBE)} \left[ \frac{\mathcal{P}_2 + \overline{\mathcal{P}}_2}{2} \right] \,, \\
\mathcal{J}^{(\PBE)}_{7} &= \frac{1}{\eps} \frac{d_{45} m_t^2 (d_{23} + d_{34})^2 -2 d_{13} d_{15} d_{23} d_{24}}{d_{15}r_{2}} \frac{\partial \mathcal{J}^{(\PBE)}_{11}}{\partial d_{34}} + \ldots   \,, \\
\mathcal{J}^{(\PBE)}_{8} &= \eps^4 r_1 d_{15} \, I_{11101111000}^{(\PBE)}\left[\frac{\mu_{12}}{(k_1 - p_{15})^2}\right] \,, \\
\mathcal{J}^{(\PBE)}_{9} &= \eps^4 r_1 \, I_{11101111000}^{(\PBE)}[\mu_{12}] \,, \\
\mathcal{J}^{(\PBE)}_{10} &= \eps^4 d_{15} d_{23} \, I_{11101111000}^{(\PBE)} \left[ (k_2 + p_1)^2 \right] \,, \\
\mathcal{J}^{(\PBE)}_{11} &= \eps^4 r_2 d_{15} \, I_{11101111000}^{(\PBE)} \,,
\end{split}
\end{equation}
where the ellipsis in the definition of $\mathcal{J}^{(\PBE)}_{7}$ denotes lengthy contributions involving the remaining MIs of this and lower sectors, whose explicit expressions are provided in the ancillary files~\cite{ancillary}.
The MIs $\mathcal{J}^{(\PBE)}_{8}$--$\mathcal{J}^{(\PBE)}_{11}$ are constructed by a combination of integrand analysis and educated guesses based on similar sectors already studied in the literature for different mass configurations.
The first two MIs are constructed by the procedure outlined in ref.~\cite{Becchetti:2025qlu}, despite the more complicated integrand, and are linked to the nested square roots.
Indeed, $\eps$-factorisation in all connection-matrix entries~---~except for those that couple to elliptic MIs, which remain linear in $\eps$~---~is achieved by the following basis change,
\begin{equation}
\label{eq:rotation_nested_PBE}
\begin{pmatrix} \mathcal{J}^{(\PBE)}_{5}  \\ \mathcal{J}^{(\PBE)}_{6} \end{pmatrix}  \mapsto  \frac{d_{45}}{a_1'} \begin{pmatrix} n_2^+ & 0 \\ 0 & n_2^- \end{pmatrix} \cdot \begin{pmatrix} 1 & 1 \\  -1 & 1 \end{pmatrix} \cdot \begin{pmatrix} \mathcal{J}^{(\PBE)}_{5}  \\ \mathcal{J}^{(\PBE)}_{6} \end{pmatrix} \, ,
\end{equation}
where the doublet of nested square roots is given by
\begin{equation}
\label{eq:nested_root_PBF}
n_2^{\pm}=\frac{\sqrt{d_{23}(a_2' \mp d_{15} r_1)(a_2' d_{23}- 2 a_1' m_t^2 \pm d_{15} d_{23} r_1)}}{\pm r_1} \,.
\end{equation}
The division by $d_{45}$ in the definition of $\mathcal{J}^{(\PBE)}_{5}$ and $\mathcal{J}^{(\PBE)}_{6}$ is arbitrary and ensures that all MIs have the same scaling dimension in $\vec{x}$.
The remaining MI, $\mathcal{J}^{(\PBE)}_{7}$, was more challenging to construct, and this is reflected in the more complicated definition.
To obtain it, we tried out various derivatives of other MIs to achieve linear $\eps$-dependence.
We then changed variables from $\vec{x}$ to momentum-twistor variables~\cite{Hodges:2009hk,Badger:2016uuq} to rationalise the square root $r_1$, applied the Magnus exponential method to integrate out the $\eps^0$ terms of the relevant connection-matrix entries, and finally changed variables back to $\vec{x}$.

Before closing this section, we make a few general remarks concerning the nested square roots discussed above.
First, since the ultraviolet and infrared poles of two-loop amplitudes are fully determined by tree-level and one-loop information, nested square roots can only appear from order $\eps^0$ onwards.
This property, however, need not be manifest in an arbitrary basis of MIs.
Using numerical evaluations with \textsc{AMFlow}~\cite{Liu:2017jxz, Liu:2021wks, Liu:2022chg, Huang:2026rjb}, we verified that, in the bases given above, the MIs whose leading singularities contain nested square roots vanish through order $\eps^3$.
We also verified explicitly that the coefficients of these MIs in the two-loop leading-colour amplitudes for $t\bar{t}\gamma$ production are finite as $\eps\to 0$.
These MIs therefore do not contribute to the poles of the two-loop amplitudes, as was observed for $t\bar{t}+$jet~\cite{Badger:2024dxo,Badger:2025ljy} and $t\bar{t}W$~\cite{Becchetti:2026awn} production, where the bases were constructed in a similar way.
This property plays an important role in the solution in terms of special functions discussed in \cref{sec:Solution}.
Second, regarding the singular locus of the nested square roots, we find that the products $r_1^4 n_1^+ n_1^-$ and $r_1^4 n_2^+ n_2^-$ factorise into polynomials in the kinematic variables $\vec{x}$, all of which belong to the alphabet of the DEs discussed in \cref{sec:DEs}. 
Finally, the outer square root of the nested square roots $n_{i}^{\pm}$ is rationalised in the limit $m_t^2 \to 0$.

\subsection{Elliptic curves}
\label{sec:elliptics}

\begin{figure}[t]
\centering
\begin{subfigure}[b]{0.497\textwidth}
\centering
\includegraphics[scale=0.58]{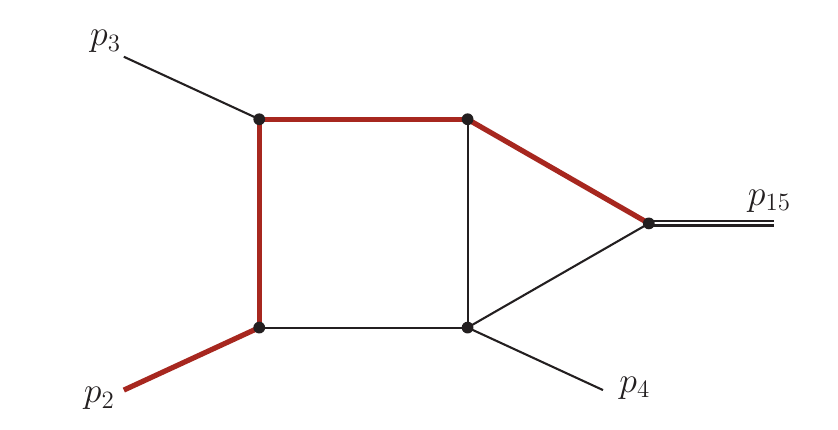}
\caption{Sector $I_{10101111000}^{(\PBE)}$ (MIs \#35--38)}
\end{subfigure}
\hfill
\begin{subfigure}[b]{0.496\textwidth}
\centering
\includegraphics[scale=0.58]{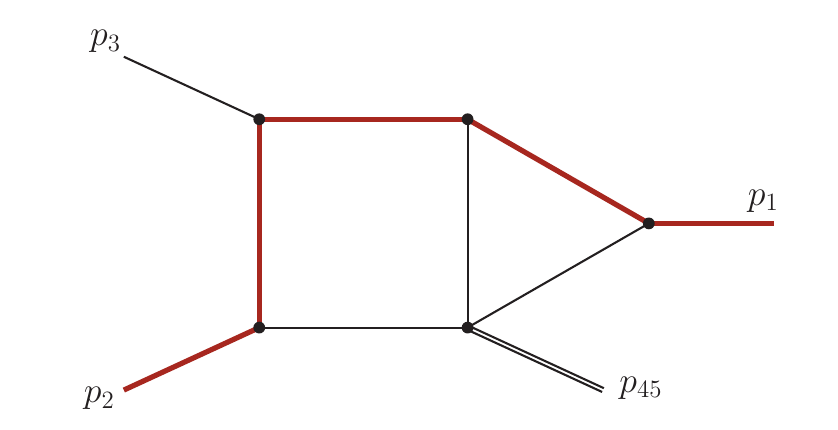}
\caption{Sector $I_{11001111000}^{(\PBE)}$ (MIs \#39--41)}
\end{subfigure}
\hfill
\begin{subfigure}[b]{0.496\textwidth}
\centering
\includegraphics[scale=0.58]{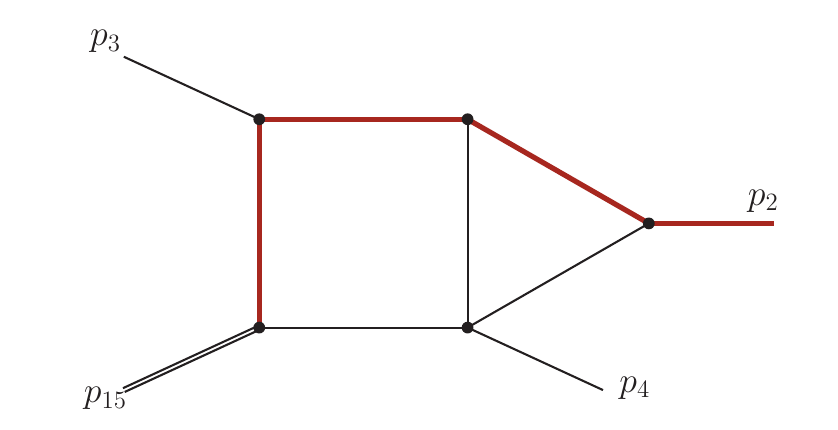}
\caption{Sector $I_{11011101000}^{(\PBF)}$ (MIs \#39--41)}
\end{subfigure}
\caption{Independent elliptic sectors appearing in the integral families under study and, in parentheses, the position of the corresponding MIs in our bases.  \PBF\ has two additional elliptic sectors, related to those shown here by a permutation of the external momenta. }
\label{fig:elliptic_sectors}
\end{figure}

There are five elliptic sectors, which do not couple to each other:
\begin{align*}
I_{11001111000}^{(\PBE)}\,, \quad I_{10101111000}^{(\PBE)}\,, \quad I_{11011101000}^{(\PBF)} \,, \quad I_{11101101000}^{(\PBF)} \,, \quad I_{11101011000}^{(\PBF)} \,.
\end{align*}
All of them have four-point kinematics and a box-triangle topology with three massive propagators, but only three of them are independent up to permutations of the external momenta.
Three independent sectors are shown in \cref{fig:elliptic_sectors}. 
The sector $I_{11101101000}^{(\PBF)}$ (MIs \#42-44) is related to $I_{11001111000}^{(\PBE)}$ by the exchange $p_1 \leftrightarrow p_2$, while $I_{11101011000}^{(\PBF)}$ (MIs \#32-35) is obtained from $I_{10101111000}^{(\PBE)}$ by exchanging $p_1 \leftrightarrow p_2$ and $p_4 \leftrightarrow p_5$.
These sectors are closely related to the box-triangle sectors studied in refs.~\cite{Badger:2024fgb, Becchetti:2025qlu}, differing only in the presence of an additional massive propagator. 
As discussed in \cref{sec:DEs}, we do not attempt to construct canonical DEs for these sectors, and instead follow the strategy adopted in refs.~\cite{Badger:2024fgb, Becchetti:2025qlu}.

To identify the elliptic curves associated with these sectors, we analysed their corner integrals in the loop-by-loop Baikov representation on the maximal cut, denoted by the superscript `$\text{M.C.}$'. 
Setting $\eps=0$ and neglecting overall kinematics-independent factors, we find that all three sectors share the generic structure
\begin{equation}
\label{eq:elliptic_Baikov}
\mathcal{J}^{\text{elliptic}}_i \bigr|_{\eps \to 0}^{\text{M.C.}} \propto \int \frac{\dd z}{\sqrt{y_i(z;\vec{x})}} \, ,
\end{equation}
where $\mathcal{J}^{\text{elliptic}}_i$, with $i\in\{a,b,c\}$, denotes the corner integral of the elliptic sector shown in \cref{fig:elliptic_sectors}~(i),
and $y_i(z;\vec{x})$ is a degree-4 polynomial in the remaining Baikov variable $z$:
\begin{equation}
\label{elliptic_curves}
\begin{split}
y_a(z;\vec{x}) &= z (z - 4m_t^2) d_{23} \left[ 4d_{13} m_t^2 (2d_{45}-z) + d_{23} (4d_{13}^2 - 4 (d_{13} + m_t^2) z + z^2)\right] \,, \\
y_b(z;\vec{x}) &= z(z-4m_t^2)\left\{ \left[2d_{15}d_{23}+(d_{23}+d_{34})(z-2d_{23})\right]^2-4d_{34}(d_{23}+d_{34})z m_t^2\right\} \,, \\
y_c(z;\vec{x}) &= \left[4z m_t^2-(z - 2d_{15})^2\right] d_{23} \left[ d_{23} (z - 2d_{24})^2 + 4 z d_{34} m_t^2 \right] \,.
\end{split}
\end{equation}
For generic $\vec{x}$, the $y_i(z;\vec{x})$'s have distinct roots and hence define an elliptic curve each.
We verified that the curves have different $j$-invariants, and thus that they are not related by isomorphisms.
Furthermore, the discriminant of each curve vanishes in the limit $m_t^2 \to 0$, as the four branch points degenerate into double roots, dropping the genus of the curve to zero. 
This observation is in agreement with the fact that the corresponding massless integrals are known to be polylogarithmic~\cite{Gehrmann:2015bfy,Papadopoulos:2015jft}.

Two of the independent sectors, $I^{(\PBE)}_{11001111000}$ and $I^{(\PBF)}_{11011101000}$, contain three MIs each. 
For each of these sectors, we construct MIs such that the diagonal block of the connection matrix corresponding to the relevant maximal cut takes the schematic form
\begin{equation} \label{eq:ellDEform3}
 \begin{pmatrix} \ast + \ast \eps & \ast & \ast \eps \\ (2\eps - 1) (\ast+\ast \eps) \, \, & \ast+  \ast \eps \, \, &  (2\eps - 1) \ast \eps\\  \ast + \ast \eps & \ast & \ast \eps  \end{pmatrix} \,,
\end{equation}
where each asterisk denotes a distinct non-zero term. 
This block can be brought to a linear dependence in $\eps$ by dividing the second MI of the sector by $2 \eps-1$. 
However, this transformation would introduce factors of $2 \eps-1$ in the denominators of the subsector entries of the connection matrix, which we prefer to avoid in order to employ the method of ref.~\cite{Badger:2024dxo} to solve the DEs (see \cref{sec:Solution}).
For the sector $I^{(\PBE)}_{11001111000}$, the form in \cref{eq:ellDEform3} is obtained by choosing the following MIs,
\begin{equation}
\label{eq:elliptic_1}
\begin{split}
& \mathcal{J}^{(\PBE)}_{39} = \eps^4 d_{23} I^{(\PBE)}_{11001111000} \left[2 \, k_2 \cdot p_1 \right]\,, \\
& \mathcal{J}^{(\PBE)}_{40} = \eps^4 (1-2\eps) d_{23} I^{(\PBE)}_{11001111000} \left[2 \, k_1 \cdot p_3 \right] \,, \\
& \mathcal{J}^{(\PBE)}_{41} = \eps^4 d_{23} (d_{12}+d_{23}-d_{45}+m_t^2) \, I^{(\PBE)}_{11001111000} \,,
\end{split}
\end{equation}
while for $I^{(\PBF)}_{11011101000}$ it is obtained with
\begin{equation}
\label{eq:elliptic_2}
\begin{split}
& \mathcal{J}^{(\PBF)}_{39} = \eps^4 (d_{23}+d_{34})   I^{(\PBF)}_{11011101000} \left[2 \, k_1 \cdot p_2 \right] \,, \\
& \mathcal{J}^{(\PBF)}_{40} = \eps^4 (1-2\eps) (d_{23}+d_{34}-d_{15}) I^{(\PBF)}_{11011101000} \left[2 \, k_2 \cdot p_3  \right] \,, \\
& \mathcal{J}^{(\PBF)}_{41} = \eps^4 d_{23} (d_{23}+d_{34}-d_{15}) \, I^{(\PBF)}_{11011101000} \,.
\end{split}
\end{equation}

The remaining independent elliptic sector, $I^{(\PBE)}_{10101111000}$, has four MIs. 
In this case, the connection matrix on the maximal cut has the form
\begin{equation}
\begin{pmatrix} \ast \eps & \ast \eps & \ast & \ast + \ast \eps \\ \ast \eps  & \ast \eps &  \ast & \ast +  \ast \eps \\  \eps (\ast + \ast \eps) & \eps (\ast +\ast \eps) & \ast+ \ast \eps & \ast + \ast \eps + \ast \eps^2 \\ \ast \eps & \ast \eps & \ast & \ast + \ast \eps \end{pmatrix} \, ,
\end{equation}
with the following basis:
\begin{equation}
\label{eq:elliptic_3}
\begin{split}
& \mathcal{J}^{(\PBE)}_{35} = \eps^3 d_{23} (d_{23}+d_{34}-d_{15}) m_t^2 \, I^{(\PBE)}_{20101111000} \,, \\
& \mathcal{J}^{(\PBE)}_{36} = \eps^4 d_{23} I^{(\PBE)}_{10101111000} \left[ 2 \, k_2 \cdot p_{234} \right] \,, \\
& \mathcal{J}^{(\PBE)}_{37} = \eps^4 (1-2\eps) d_{23} I^{(\PBE)}_{10101111000} \left[ 2 \, k_1 \cdot p_3 \right] \,, \\
& \mathcal{J}^{(\PBE)}_{38} = \eps^4 d_{23} (d_{23}+d_{34}-d_{15}) \, I^{(\PBE)}_{10101111000} \,.
\end{split}
\end{equation}
The last three integrals in \cref{eq:elliptic_3} together with the ones of \cref{eq:elliptic_1,eq:elliptic_2} were chosen following the strategy of refs.~\cite{Badger:2024fgb,Becchetti:2025qlu}.
Their normalisation is chosen arbitrarily so that all MIs have the same scaling dimensions.
$\mathcal{J}^{(\PBE)}_{35}$, on the other hand, was constructed through integrand analysis in the Baikov representation on the maximal cut.

Numerical evaluations with \textsc{AMFlow} show that all MIs in \cref{eq:elliptic_1,eq:elliptic_2,eq:elliptic_3} vanish through $\mathcal{O}(\eps^3)$, except for $\mathcal{J}^{(\PBE)}_{35}$, which is non-zero starting at that order.
However, the solution method of \cref{sec:Solution} reveals that the $\mathcal{O}(\eps^3)$ term of $\mathcal{J}^{(\PBE)}_{35}$ is an iterated integral of logarithmic one-forms, and is therefore free of elliptic functions.
As we argued for the nested square roots at the end of the previous section, and in line with refs.~\cite{Badger:2024dxo,Badger:2025ljy,Becchetti:2026awn}, we will see in \cref{sec:Solution} that these properties allow us to obtain a representation of the two-loop leading-colour amplitudes for $t\bar{t}\gamma$ production in which the poles in $\eps$ are manifestly free of elliptic functions.
Furthermore, given the similarity between the bases introduced here and those of refs.~\cite{Badger:2024fgb, Becchetti:2025qlu}, we expect the former to be a suitable starting point for obtaining canonical DEs, as done in ref.~\cite{Becchetti:2025oyb} for \PBB.

\section{Special functions}
\label{sec:Solution}

In this section, we derive a representation of the integral bases constructed above that facilitates their numerical evaluation and the computation of the relevant amplitude.
To this end, we are interested in the expansion of the integral bases around $\eps=0$:
\begin{align}
\vec{\cal J}^{({\rm F})}(\vec{x}, \eps) = \sum_{k=0}^{4} \eps^k \, \vec{\cal J}^{({\rm F}, k)}(\vec{x}) + \mathcal{O}\left(\eps^5\right) \,.
\end{align}
The expansion starts at $\eps^0$ as a consequence of the normalisation of the bases.
We have explicitly verified that terms up to order $\eps^4$ are sufficient to compute the two-loop amplitude through order $\eps^0$, as required for NNLO calculations.\footnote{We are grateful to Heribertus Bayu Hartanto for his help with this check.}
While this alignment between the $\eps$ expansion of amplitudes and master integrals~---~i.e.\ that the same number of terms is needed in both expansions~---~is well established in the case of canonical bases, higher orders in $\eps$ are generally needed for arbitrary bases.
The fact that our bases can be truncated at order $\eps^4$ is therefore a non-trivial check that they have the desired properties, and is consistent with the findings of other applications of this approach to integral bases~\cite{Badger:2024dxo,Becchetti:2025qlu,Pozzoli:2026eiu}.

In order to obtain boundary conditions to solve the DEs and perform checks, we evaluate the bases of all families at 12 arbitrary phase-space points in the physical $s_{45}$ channel.
To this end, we employ \textsc{AMFlow} interfaced to \textsc{LiteRed} and \textsc{FiniteFlow}, and request a precision of 70 digits. 
The values can be found in our ancillary files~\cite{ancillary}.

While the master integrals $\vec{\cal J}^{({\rm F})}(\vec{x},\eps)$ are linearly independent by construction, the coefficients $\vec{\cal J}^{({\rm F}, k)}(\vec{x})$ in their $\eps$ expansion are generally redundant.
An expression of the amplitude in terms of these coefficients therefore contains hidden zeroes, which complicate the resulting expressions, prevent the exact subtraction of ultraviolet and infrared poles required to obtain the finite remainder, and lead to numerical cancellations and loss of precision.
A powerful approach to overcome these issues is to express the master-integral coefficients $\vec{\cal J}^{({\rm F}, k)}(\vec{x})$ as polynomials in a \emph{generating set of algebraically independent special functions} $f^{({\rm F},k)}_{j}(\vec{x})$ and transcendental constants, with constant rational coefficients. 
For example, we aim to write
\begin{align} \label{eq:exJ}
\mathcal{J}^{(\PBD, 2)}_{91} = 
 3 \, \zeta_2 + 
  2 \, \Bigl(f^{(\PBD,1)}_{1}\Bigr)^2 + 4 \, f^{(\PBD, 1)}_{1} f^{(\PBD, 1)}_{3} - 4 \, \Bigl( f^{(\PBD, 1)}_{3}\Bigr)^2 +
    2 \, f^{(\PBD, 2)}_{7} \,,
\end{align}
where $\zeta_n$ is the Riemann zeta function and we omitted the argument $\vec{x}$.
In this work, we construct a generating set of special functions separately for each family. 
A common generating set is advantageous for the amplitude calculation, as it enables cancellations across different families and crossings.
Indeed, we have already seen in \cref{sec:Bases_and_DEs} that the families share common subsectors.
Here, however, we find it more convenient to apply the numerical evaluation method discussed in \cref{sec:numerics} separately to each family (see also ref.~\cite{Becchetti:2026yxl}).
We defer to future work the merging of all families and crossings, including those contributing to $t\bar{t}+$jet production, for which a generating set of special functions is already available~\cite{Badger:2024fgb}.

When the integral bases satisfy canonical DEs, they can be expressed in terms of Chen iterated integrals~\cite{Chen:1977oja}, whose properties enable an algorithmic construction of a representation in terms of independent special functions~\cite{Gehrmann:2018yef,Chicherin:2020oor}.
In this case, the superscript $k$ in both $\mathcal{J}^{({\rm F}, k)}_{i}$ and $f^{({\rm F}, k)}_{i}$ denotes the transcendental weight of the function, namely the depth of its iterated integral representation.
The basis functions $f^{({\rm F},k)}_{i}(\vec{x})$ may be selected from the master-integral coefficients themselves and graded by their analytic properties~\cite{Chicherin:2021dyp}, such as their branch-cut structure, in such a way that only $\zeta_2$ and $\zeta_3$ are expected to appear as transcendental constants up to the $\eps$ order considered here~\cite{Abreu:2023rco}.
Since the basis of \PBD\ satisfies canonical DEs, we straightforwardly apply the algorithm of ref.~\cite{Abreu:2023rco}, resulting e.g.\ in  \cref{eq:exJ}, with the appearing functions defined in terms of master-integral coefficients as
\begin{align}
f^{(\PBD,1)}_{1} (\vec{x}) & =  - \frac{1}{2} \mathcal{J}^{(\PBD,1)}_{100}(\vec{x}) \,, \\
f^{(\PBD,1)}_{3}(\vec{x}) & = \frac{1}{2}  \mathcal{J}^{(\PBD,1)}_{91}(\vec{x}) - \frac{1}{2}  \mathcal{J}^{(\PBD,1)}_{100}(\vec{x}) \,, \\
f^{(\PBD,2)}_{7}(\vec{x}) & =   \mathcal{J}^{(\PBD,2)}_{90}(\vec{x}) \,.
\end{align}

For the families \PBE\ and \PBF, instead, we do not have canonical DEs.
Therefore, we cannot write their bases straightforwardly in terms of iterated integrals and apply the same method as for \PBD.
However, as discussed in \cref{sec:Bases_and_DEs}, we constructed the bases for these two families following the criteria of ref.~\cite{Badger:2024dxo}.
As a result, while their DEs take a non-canonical form (cf.\ \cref{eq:DEs_general_2,eq:our_connection_matrices}), we verified by numerical evaluation that they satisfy the following constraints:
\begin{align} \label{eq:non_canonical_constraints}
\begin{aligned}
& \alpha_j^{(\mathrm{F},0)} \cdot \vec{\mathcal{J}}^{(\mathrm{F},k)}(\vec{x}) = \beta_j^{(\mathrm{F},0)} \cdot \vec{\mathcal{J}}^{(\mathrm{F},k)}(\vec{x}) = 0  \,, \\
& \beta_j^{(\mathrm{F},1)} \cdot \vec{\mathcal{J}}^{(\mathrm{F},k-1)}(\vec{x}) = 0 \,, \\
& \alpha_j^{(\mathrm{F},2)} \cdot \vec{\mathcal{J}}^{(\mathrm{F},k-2)}(\vec{x}) = \beta_j^{(\mathrm{F},2)} \cdot \vec{\mathcal{J}}^{(\mathrm{F},k-2)}(\vec{x}) = 0  \,, \\
\end{aligned}
\end{align}
for $k=0,\ldots,3$, where $\vec{\mathcal{J}}^{(\mathrm{F},k)}(\vec{x}) \equiv 0$ for $k<0$.
Effectively, these remove all non-canonical features from the DEs for the master-integral coefficients $\vec{\cal J}^{({\rm F}, k)}(\vec{x})$ with $k$ up to $3$, which can thus be treated similarly to \PBD.
Elliptic curves and nested square roots are bound to appear at $k=4$, the highest order we need, but affect only a small subset of the master-integral coefficients, which we denote by $S_{\rm F}$.
Concretely, this means that the elements of the constraints in \cref{eq:non_canonical_constraints} that do not belong to $S_{\rm F}$ hold also for $k=4$.
E.g.\ we have that
\begin{align}
\bigl[\beta_j^{(\mathrm{F},1)} \cdot \vec{\mathcal{J}}^{(\mathrm{F},3)}(\vec{x})  \bigr]_i = 0 \quad \forall \, i \notin S_{\rm F} \,,
\end{align}
and similarly for the other terms in \cref{eq:non_canonical_constraints}.
Therefore, we can treat ${\cal J}_i^{({\rm F}, 4)}(\vec{x})$ for $i\notin S_{\rm F}$ as above.

Finally, we are left with the master-integral coefficients ${\cal J}_i^{({\rm F}, 4)}(\vec{x})$ with $i\in S_{\rm F}$ for \PBE\ and \PBF.
Explicitly, these sets are given by
\begin{align}
S_{\PBE} & = \bigl\{5, 6, 7, 12, 13, 20, 35, 36, 37, 38, 39, 40, 41 \bigr\} \,, \\
S_{\PBF} & = \bigl\{4, 5, 6, 15, 17, 18, 32, 33, 34, 35, 39, 40, 41,42, 43, 44  \bigr\} \,.
\end{align}
For \PBE, the set includes the master integrals of the two elliptic sectors (35--38 and 39--41) and three of the nested-square-root sector (5,6,7).
The remaining three master integrals (12, 13 and 20) belong to logarithmic sectors, but their derivatives are coupled to the elliptic sectors at order $\eps^4$.
For \PBF, the set includes master integrals of the three elliptic sectors (32--35, 39--41 and 42--44), five of the master integrals of the two nested-square-root sectors (4--10 and 17--19), and a single master integral (15) which belongs to a logarithmic sector but is coupled to the master integrals of an elliptic sector at order $\eps^4$.
Following the approach of ref.~\cite{Badger:2024dxo}, we add the master-integral coefficients for which we lack an iterated-integral representation to the generating set, and denote them by $f^{(\mathrm{F},4^*)}_{i} (\vec{x})$.

In conclusion, for each family, we have represented the basis order by order in $\eps$ through $\eps^4$ as polynomials in a generating set of special functions ($f^{(\mathrm{F},k)}_{i} (\vec{x})$) and transcendental constants ($\zeta_2$ and $\zeta_3$).
The relevant expressions can be found in our ancillary files~\cite{ancillary}.
For illustration, we give here a few representative terms in the expression of $\mathcal{J}^{(\PBE)}_{20}$:
\begin{align}
\begin{aligned}
\mathcal{J}^{(\PBE)}_{20} & = \frac{\eps}{24} \left( f^{(\PBE, 1)}_{2} - f^{(\PBE, 1)}_{4} \right) + \frac{\eps^2}{48} \biggl[ 5 \, f^{(\PBE,2)}_9 + 4 \, f^{(\PBE,1)}_1 f^{(\PBE,1)}_2 - 15 \, \zeta_2 + \cdots \biggr] \\
& \phantom{=} \ + \frac{\eps^3}{5184} \biggl[ 2496 \, f^{(\PBE, 3)}_{52} +3456 \, f^{(\PBE, 1)}_{3} f^{(\PBE, 2)}_{7} - 11393 \, \zeta_3 + \cdots \biggr] \\
& \phantom{=} \ + \eps^4 \, f^{(\PBE, 4^*)}_{6} + \mathcal{O}\left(\eps^5\right) \,,
\end{aligned}
\end{align}
where we omitted the arguments and the ellipses denote polynomials in the special functions with homogeneous transcendental weight.
We give the number of functions in the generating set for each family in \cref{tab:functions}.
By construction, the functions in the generating set inherit from the corresponding master integrals a definite charge (even or odd) with respect to each square root (cf.\ \cref{sec:Bases_and_DEs}).

\begin{table}[h]
\begin{center}
\begin{tabular}{c|c|c|c|c|c|c|c}
family & MI & $f^{(\mathrm{F},1)}$ & $f^{(\mathrm{F},2)}$ & $f^{(\mathrm{F},3)}$ & $f^{(\mathrm{F},4)}$ & $f^{(\mathrm{F},4^*)}$ & all $f^{(\mathrm{F},k)}$  \\
\hline
\PBD  & 100 & 6 & 10 & 46 & 85 & 0 & 147 \\ %& 177 \\
\PBE  & 154 & 6 & 10 & 53 & 110 & 13 & 192 \\ % & 240 \\
\PBF  & 142 & 6 & 10 & 51 & 102 & 16 & 185 \\ %& 242 \\
\end{tabular}
\end{center}
\caption{Number of master integrals (MI) and special functions in the generating set for each family.
%DE basis refers to the closure of the function set upon differentiation, which requires the inclusion of a number of polynomials as discussed at the end of \cref{sec:Solution}.
}
\label{tab:functions}
\end{table}

The functions $f^{({\rm F}, k)}_i$ with $k=1,2,3,4$ admit a representation as weight-$k$ iterated integrals of logarithmic one-forms and are algebraically independent.
The weight-1 functions play a special role, as they appear not only in the bare amplitudes through Feynman integrals, but also explicitly as logarithms in the IR subtraction terms.
Therefore, we choose them to be the same across the three families, and write them here in terms of logarithms in the $s_{45}$ channel:
\begin{equation}
\begin{aligned}
f^{({\rm F},1)}_{1} & = \log(m_t^2) \,, &
f^{({\rm F},1)}_{2} & = \log(-2 d_{15}) \,,  &
f^{({\rm F},1)}_{3} & = \log(-2 d_{24}) \,, \\
f^{({\rm F},1)}_{4} & = \log(2 d_{23}) - \mathrm{i} \pi \,,  &
f^{({\rm F},1)}_{5} & = \log(2 d_{45}) - \mathrm{i} \pi \,,  &
f^{({\rm F},1)}_{6} & = \log(2 d_{13}) -\mathrm{i} \pi \,.
\end{aligned}
\end{equation}
Note that we refrain from grading the functions $f^{({\rm F}, k)}_i$ by their analytic properties as done in ref.~\cite{Chicherin:2021dyp}, since at the level of the amplitude they will mix with the functions $f^{({\rm F}, 4^*)}_i$, whose grading is impossible owing to the lack of an iterated-integral representation.

For the functions $f^{({\rm F}, 4^*)}_i$, instead, no such representation is currently available, and the label $4^*$ merely indicates that these functions first appear at order $\eps^4$ in the expansion of the corresponding integrals.
We have verified with the PSLQ algorithm~\cite{pslq} that they do not satisfy $\mathbb{Q}$-linear relations, but we cannot exclude
 the existence of more complicated relations.
However, the functions $f^{({\rm F}, 4^*)}_i$ are relatively few and the fact that they are associated with distinct analytic features~---~i.e.\ different elliptic curves and nested square roots~---~provides a heuristic indication that the number of potentially missed relations is limited.
Moreover, as we mentioned above, we have verified that the coefficients that multiply our integral bases in the two-loop amplitudes for $t\bar{t}\gamma$ production contain poles no higher than $1/\eps^4$. 
In particular, this means that the functions $f^{({\rm F}, 4^*)}_i$ can only appear in the terms of order $\eps^0$ of the two-loop amplitudes, and therefore that the IR and UV poles can be subtracted exactly.

Before moving on to the numerical evaluation, we close this section with an observation that will enable the method discussed in the next section: 
by construction, the generating set of functions $f^{({\rm F}, k)}_i$ for each family is closed under differentiation.
In other words, the derivative of any function in the set can be written as a polynomial in the functions themselves:
\begin{align} \label{eq:df}
\mathrm{d} f^{(\mathrm{F},k)}_i(\vec{x}) = \sum_j a^{(\mathrm{F},k)}_{ij} \, \mathrm{d} \log\left( W_j(\vec{x})\right) \, Q^{(\mathrm{F},a)}_j(\vec{x}) + \sum_j b^{(\mathrm{F},k)}_{ij} \, \omega_j(\vec{x}) \, Q^{(\mathrm{F},b)}_j(\vec{x}) \,,
\end{align}
where $a^{(\mathrm{F},k)}_{ij}$ and $b^{(\mathrm{F},k)}_{ij}$ are rational numbers, while $Q^{(\mathrm{F},a)}_j(\vec{x})$ and $Q^{(\mathrm{F},b)}_j(\vec{x})$ are polynomials in the special functions themselves and the transcendental constants, namely
\begin{align} \label{eq:Q}
Q^{(\mathrm{F},a)}_j(\vec{x}) \in \mathbb{Q}\left[ \left\{ f^{({\rm F},k)}_i(\vec{x}) \bigl| k\in\{1,2,3,4,4^*\}, i=1,\ldots \right\} \cup \left\{ \zeta_2, \zeta_3 \right\} \right] \,,
\end{align}
and similarly for $Q^{(\mathrm{F},b)}_j(\vec{x})$.
The non-linear DEs in \cref{eq:df} can be read off from the DEs for the integral bases, but are substantially simpler as a result of the truncation of the $\eps$ expansion. 
Not only are the DEs for the special functions independent of $\eps$, but they also involve significantly fewer one-forms, as shown in \cref{tab:diff-forms}. 
This reduction translates into a more efficient numerical solution, as discussed in the next section.
\begin{table}[h]
\begin{center}
\begin{tabular}{c|c|c}
family
& $\mathrm{d}\log$ one-forms 
& non-$\mathrm{d}\log$ one-forms \\
\hline
\PBD & $77 \to 73$ & $0 \to 0$ \\
\PBE & $113 \to 97$ & $177 \to 87$ \\
\PBF & $129 \to 107$ & $222 \to 120$
\end{tabular}\end{center}
\caption{Reduction in the number of differential forms when going from the differential equations for the master integrals to those for the special functions.}
\label{tab:diff-forms}
\end{table}

The method discussed in \cref{sec:numerics} also works with non-linear DEs, but several public tools commonly used in the field solve linear DEs by series expansions~\cite{Pozzorini:2005ff,Moriello:2019yhu}.
To allow our special functions to be evaluated with these tools as well, we recast \cref{eq:df} as a linear system by enlarging the generating set of each family. 
We iteratively add the polynomials in the special functions and transcendental constants that appear in the derivatives of the elements, until the set closes under differentiation.
At each step, we add only polynomials that are $\mathbb{Q}$-linearly independent of the elements already in the set. 
Collecting the special functions and the additional polynomials into a vector $\vec{F}^{(\mathrm{F})}(\vec{x})$, we obtain the linear DEs
\begin{align} \label{eq:linear_DEs}
\dd \vec{F}^{(\mathrm{F})}(\vec{x}) = \Omega^{(\mathrm{F})}(\vec{x}) \, \vec{F}^{(\mathrm{F})}(\vec{x}) \,,
\end{align}
where the entries of $\Omega^{(\mathrm{F})}(\vec{x})$ are $\mathbb{Q}$-linear combinations of the one-forms appearing in \cref{eq:df}.
We include a \textsc{Mathematica} script that solves the linear DEs for the special functions with \textsc{DiffExp} in our ancillary files~\cite{ancillary}.

We validated the DEs in \cref{eq:df} by solving them in the linear form of \cref{eq:linear_DEs} with \textsc{DiffExp}~\cite{Hidding:2020ytt}, using the \textsc{AMFlow} values at one of the 12 phase-space points as boundary conditions, and comparing the results at the remaining 11 points with the corresponding \textsc{AMFlow} values. We also verified that our expressions for the MIs in terms of special functions are consistent with the \textsc{AMFlow} results at all 12 points. 

\section{Numerical evaluation}
\label{sec:numerics}

In this section, we discuss the numerical evaluation of the special functions defined in \cref{sec:Solution}.
We follow the approach adopted for $t\bar{t}+$jet production in ref.~\cite{Badger:2025ljy}, which is based on numerically integrating the DEs satisfied by the special functions (cf.~\cref{eq:df}) along a path starting from one of the available starting points.
We perform the integration using the Bulirsch--Stoer algorithm (see e.g.~\cite{Hairer}), as implemented in the \textsc{odeint} library of \textsc{Boost}~\cite{BoostLibrary}, following its first application to similar problems in ref.~\cite{Czakon:2020vql}.
We refer the interested reader to refs.~\cite{Boughezal:2006px,Czakon:2008zk} for the earliest applications of this algorithm to Feynman integrals and to refs.~\cite{PetitRosas:2025xhm,Pozzoli:2026eiu,Czakon:2026tog,Gotzfried:2026def} for recent related studies.

We solve the DEs associated with the special functions of each family separately for several reasons. Combining all families and their permutations relevant to the amplitude computation would result in a very large system, making its compilation and optimisation computationally demanding. Moreover, 
spurious poles close to the integration contour can cause numerical instabilities. Treating each family separately therefore minimises the number of spurious singularities present in any given system, avoiding the accumulation of such singularities in a single, larger system. Redundancies can also be eliminated without merging the families, as done in ref.~\cite{Becchetti:2026yxl}. 

While it is possible to integrate algebraic DEs numerically~\cite{Badger:2025ljy,Czakon:2026tog}, we find it convenient to remove all square roots from the problem, so as to work with rational expressions only.
In practice, we solve the DEs satisfied by the auxiliary functions $g^{(\mathrm{F},k)}_i$, defined as
\begin{align} \label{eq:g_def}
g^{(\mathrm{F},k)}_i(\vec{x}) = \frac{f^{(\mathrm{F},k)}_i(\vec{x})}{\prod_j r_j(\vec{x})^{q_{ij}}} \,,
\end{align}
where $q_{ij}=1$ if $f^{(\mathrm{F},k)}_i$ is odd under $r_j \to -r_j$, and $q_{ij}=0$ otherwise. 
The set of functions $g^{(\mathrm{F},k)}_i$ then satisfies a system of non-linear DEs analogous to \cref{eq:df,eq:Q} (with $f$ replaced by $g$), where all one-forms and letters are rational.
We write all one-forms (including the $\dlog$'s) as in \cref{eq:omega} with $R_i = 1$.

The numerical solution requires multiple evaluations of the derivatives of the special functions along the integration path.
Optimising these evaluations is therefore essential for an efficient integration, and in practice this mainly means reducing the number of multiplications.
To this end, we change the basis of one-forms,
\begin{align} \label{eq:rotation_of}
\omega_i'(\vec{x}) = \sum_j M_{ij} \, \omega_j(\vec{x}) \,,
\end{align}
with $M_{ij} \in \mathbb{Q}$ and $\det M \neq 0$, so as to \emph{decorrelate} their poles, as proposed for the coefficients of scattering amplitudes in refs.~\cite{DeLaurentis:2023nss,DeLaurentis:2025dxw}.
We construct the matrix $M$ iteratively.
Let $P(\vec{x})$ be a polynomial appearing in the denominators of the one-forms, and let $m$ be its highest exponent.
We parametrise the kinematics on a random univariate slice, $\vec{x} = \vec{a} + \vec{b}\, t$, where $\vec{a}$ and $\vec{b}$ are vectors of random integers, and use polynomial expansion~\cite{Fontana:2023amt} to compute the leading term of each one-form around the zeros of $P(t) \equiv P(\vec{x}(t))$,
\begin{align}
\omega_j(t) = \left[ \frac{\sum_{k=0}^{\mathrm{deg} P-1} e^{(k)}_{j} \, t^k}{P(t)^m} + \mathcal{O}\left(\frac{1}{P(t)^{m-1}} \right) \right]\, \dd t \,,
\end{align}
where $e^{(k)}_j \in \mathbb{Q}$ and $\mathrm{deg} P$ is the degree of $P(t)$.
We compute the coefficients $e^{(0)}_j$ on several slices and collect them into a matrix.
Its null space defines linear combinations of the one-forms that are free of the pole $P(\vec{x})^{-m}$.
We complete these to a new basis with the simplest elements of the original basis $\{ \omega_j(\vec{x}) \}$, so that the pole $P(\vec{x})^{-m}$ is isolated in as few one-forms as possible.
We repeat this for all denominator polynomials and all possible exponents, and adopt the change of basis that yields the largest reduction in the polynomial degree of the numerators, which we take as a proxy for the complexity of the one-forms.
We then apply the same procedure separately to the two resulting subsets of the basis: the one-forms that contain the pole $P(\vec{x})^{-m}$ treated in the previous step, and those that do not.
This ensures that poles which have already been decorrelated are not reintroduced.
The iteration terminates when the attempted change of basis is the identity.
The whole procedure is carried out with finite-field methods.
Compared with refs.~\cite{DeLaurentis:2023nss,DeLaurentis:2025dxw}, our strategy differs in two respects: the search algorithm we use to construct the change of basis $M$, and the use of polynomial expansion to probe the one-forms near the singularity at $P(t)=0$ without requiring explicit knowledge of the roots of $P(t)$.
The latter is computationally advantageous, since it avoids square roots, and it is in fact essential here, because some of the polynomials have degree $5$, and their roots cannot in general be expressed in radicals.
For \PBD, the change of basis brings no significant improvement, presumably because the one-forms are $\dd\log$ forms and hence already as simple as possible.
The largest improvement is for \PBE, where the maximal numerator degree of the one-forms drops from $22$ to $17$.

We then perform the following steps to further reduce the number of multiplications required to evaluate the new one-forms.
\begin{enumerate}

\item We set $m_t^2=1$, i.e.\ we evaluate the functions at the rescaled kinematics $\vec{x}/m_t^2$.\footnote{The special functions are generally not homogeneous. If needed, the overall scale should be restored at the level of the master integrals or the amplitude.}

\item We change variables from $\vec{x}$ to a set of scalar products built from pairs of momenta that are adjacent in the topologies considered in this work, i.e.
\begin{align}
\vec{y} = \left(d_{13}, d_{23}, d_{24}, d_{45}, d_{15} \right) \,.
\end{align}
We observe that this choice makes the one-forms significantly more compact.
We postpone to future work the investigation of choosing an optimal set of variables for each one-form separately~\cite{Badger:2025ljy}, which is expected to become important when all permutations contributing to the amplitude are merged.

\item We decompose the one-form coefficients ($\omega_{ij}(\vec{x})$ in \cref{eq:omega}) into multivariate partial fractions using \textsc{MultivariateApart}~\cite{Heller:2021qkz}.
In cases where a complete decomposition exceeds a prescribed time limit, we instead use the following heuristic procedure.
We remove all possible subsets of denominator factors and repeat the partial fraction decomposition.
Finally, we compare the resulting decompositions with the original expression and retain the expression with the fewest multiplications.

\item We factorise the partial fractions and write them in terms of a global set of multiplicatively independent irreducible polynomials, $\{p_m(\vec{y})\}$.

\item Finally, we write the right-hand side of the DEs in terms of a set of $\mathbb{Q}$-linearly independent polynomials $Q_n$ in the special functions and transcendental constants (cf.~\cref{eq:df,eq:Q}).

\end{enumerate}
The evaluation of the derivatives then proceeds in four stages: the polynomials $p_m$ from the kinematics $\vec{y}$, the one-forms from the $p_m$, the special-function polynomials $Q_n$ from the current values of the functions, and finally the derivatives.
We optimise each stage separately with the \texttt{\#optimize} routine of \textsc{FORM}~\cite{Ruijl:2017dtg}, which performs common-subexpression elimination on multivariate Horner schemes, and we translate the result into \textsc{C++} code, following the structure of the library provided with ref.~\cite{Badger:2025ljy}.

Following ref.~\cite{Badger:2025ljy}, we use two types of integration path from one of the available starting points $\vec{x}_0$, where the special functions are already known, to the target point $\vec{x}$.
First, we consider the straight \emph{line} in the invariants $\vec{x}$,
\begin{align}
 \vec{x}_{\mathrm{line}}(t)=\vec{x}_{0}+t\,(\vec{x}-\vec{x}_{0}) \,,
\end{align}
for $t\in [0,1]$.
As we will see below, this type of path typically yields the fastest integration.
However, the physical region is not convex in the invariants, so a straight line may leave the physical domain even when both endpoints lie inside it.
In such cases, to avoid analytic continuation across branch-cut singularities, we adopt a \emph{physical} path, $\vec{x}_{\mathrm{phys}}(t)$, along which $d_{12}$, $d_{45}$ and three angular variables are varied linearly in the parameter $t$ (see ref.~\cite{Badger:2025ljy} for a detailed discussion).
While the physical path is generally slower to integrate, it is guaranteed to remain within the physical domain.
Another complication arises from the fact that the DEs have spurious singularities on subvarieties of the phase space, which prevent a stable integration.
To avoid them, we integrate along a complex deformation of the paths $\vec{x}_{\gamma}(t)$ introduced above,
\begin{align} \label{eq:path}
x_{\gamma,j}(t) = x_{0,j} + \bigl[1 + 4\,\mathrm{i}\,\delta_j\,(1-t)\bigr]\bigl(x_{\gamma_{\text{R}},j}(t)-x_{0,j}\bigr) \,, 
\end{align}
where $\gamma_{\text{R}}$ is the real path, line or physical, before the deformation, while $\gamma$ is the path after the complex deformation.
As in ref.~\cite{Badger:2025ljy}, we deform only the positive invariants $d_{12}$, $d_{23}$ and $d_{45}$, with $\delta_j = 0.1, 0.15, 0.2$, respectively.
Note that the path is defined in terms of the invariants $\vec{x}$, and then mapped onto the variables $\vec{y}$ in which the DEs are written.

Since the amplitudes for $t\bar{t}\gamma$ production are not yet available and they are not the focus of this work, we cannot adopt the amplitude-based error estimate of ref.~\cite{Badger:2025ljy}, and we control the error on the special functions directly.
At each step, the Bulirsch--Stoer algorithm provides an estimate $\delta g_i$ of the local error on each function, obtained from the difference between successive orders of the Richardson extrapolation.
The step is accepted~if
\begin{align}
\max_i |\delta g_i| \le \epsilon_{\rm abs} \,,
\end{align}
where $\epsilon_{\rm abs}$ is the absolute tolerance, and the step size is reduced otherwise.

We tested our implementation on 11\,056 phase-space points, namely the
unweighted Born events generated in the soft fiducial phase space used for
the $pp \to t\bar{t}j$ cross-section calculation of ref.~\cite{Badger:2025ilt}.
We set $\epsilon_{\rm abs}=10^{-10}$, compile with \texttt{GCC 13.3.0} (flags \texttt{-std=c++17 -O2}), 
and run in double precision on a single core of an Intel Core i7-10750H laptop processor.
We start the integration from the arbitrary point
\begin{align*}
\left\{d_{12} = \frac{325499}{89787}, d_{23} = \frac{309277}{59858}, d_{34} = -\frac{285382}{89787},
       d_{45} = \frac{405000}{29929}, d_{15} = -\frac{532037}{89787}, m_t = 1 \right\},
\end{align*}
which corresponds to the boundary values \texttt{PB*\_X12.m} in the ancillary files~\cite{ancillary}. 
We repeated the test starting from a different boundary point, given in \texttt{PB*\_X11.m} in the ancillary files, and obtained similar results, which we omit for brevity.
We estimated the number of correct digits from the largest relative difference, over all special functions, between the results of the integration from the two boundary points, $\delta_{\max}$, as $-\log_{10}\delta_{\max}$.

\begin{figure}[t]
\centering
\includegraphics[width=0.48\textwidth]{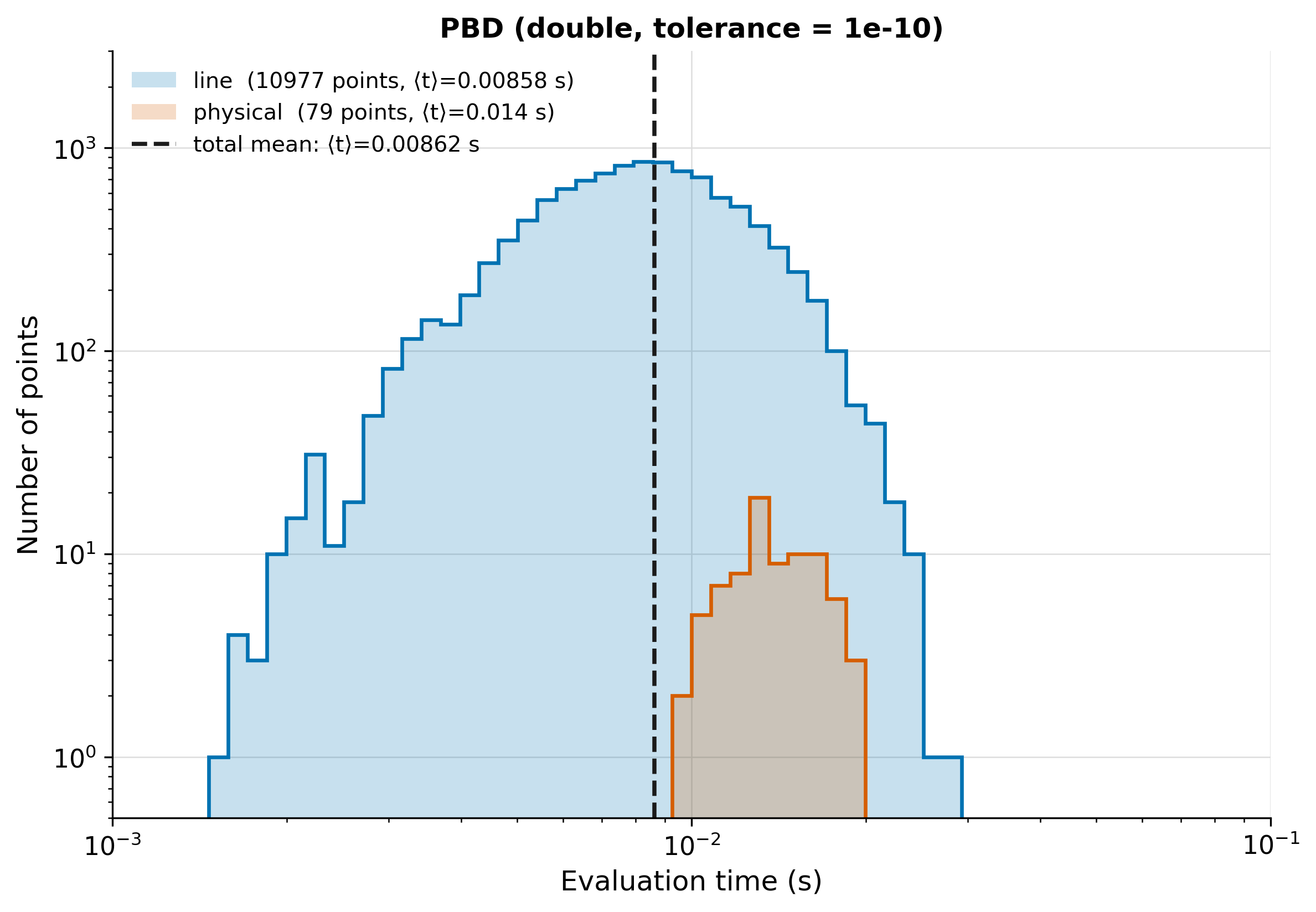}\hfill
\includegraphics[width=0.48\textwidth]{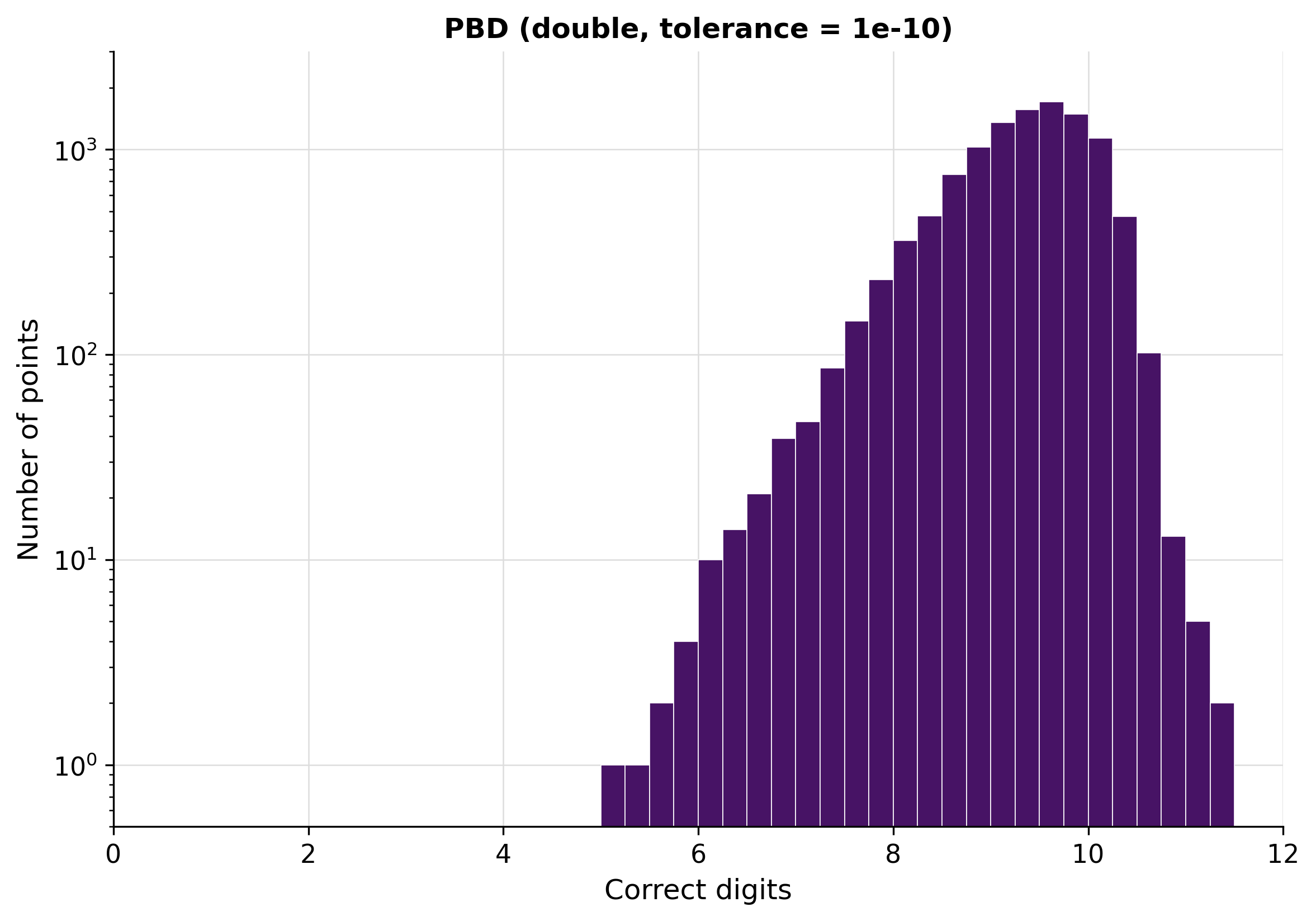}
\caption{Evaluation of the special functions of \PBD\ in double precision with $\epsilon_{\rm abs}=10^{-10}$. 
Left: distribution of the evaluation time per point, separately for the straight-line and physical paths. 
Right: distribution of the correct digits.}
\label{fig:PBD}
\end{figure}

The results for \PBD\ are shown in \cref{fig:PBD}. Since this family contains
no elliptic or nested-square-root sectors, the evaluation is fast, taking
$\mathcal{O}(10)$\,ms per point. 
The change of one-form basis in \cref{eq:rotation_of} leads to a modest $3\%$ reduction in the mean integration time.
The bulk of the points reach 9--10 correct
digits, and all of them at least 5 digits of accuracy. For fewer than $1\%$ of the points a
physical path is required, which is on average slower than the straight-line
path. 
This suggests that choosing starting points tailored to the target phase-space distribution could further speed up the evaluation, by reducing the reliance on the physical path parametrisation. 

\begin{figure}[t]
\centering
\includegraphics[width=0.48\textwidth]{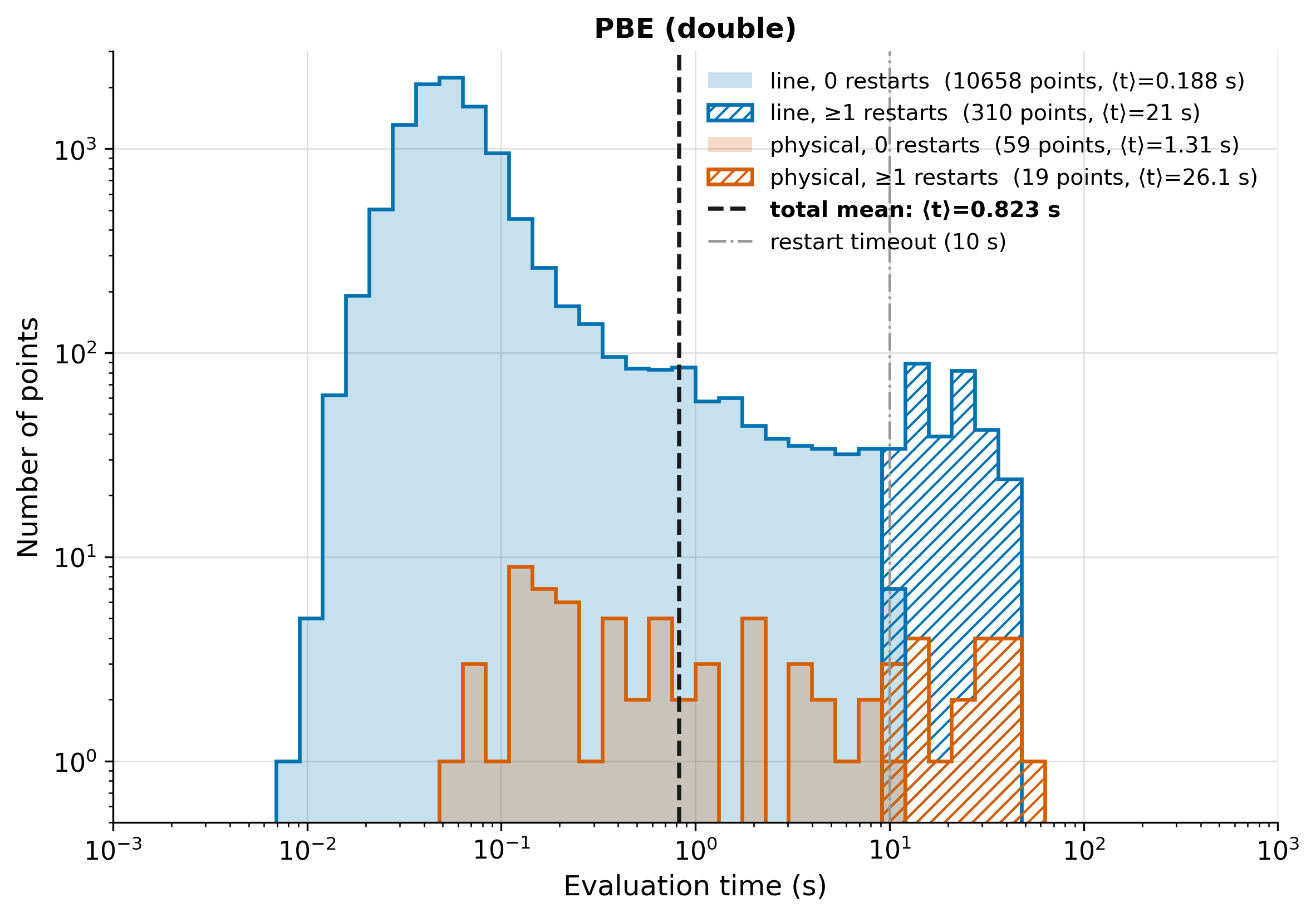}\hfill
\includegraphics[width=0.48\textwidth]{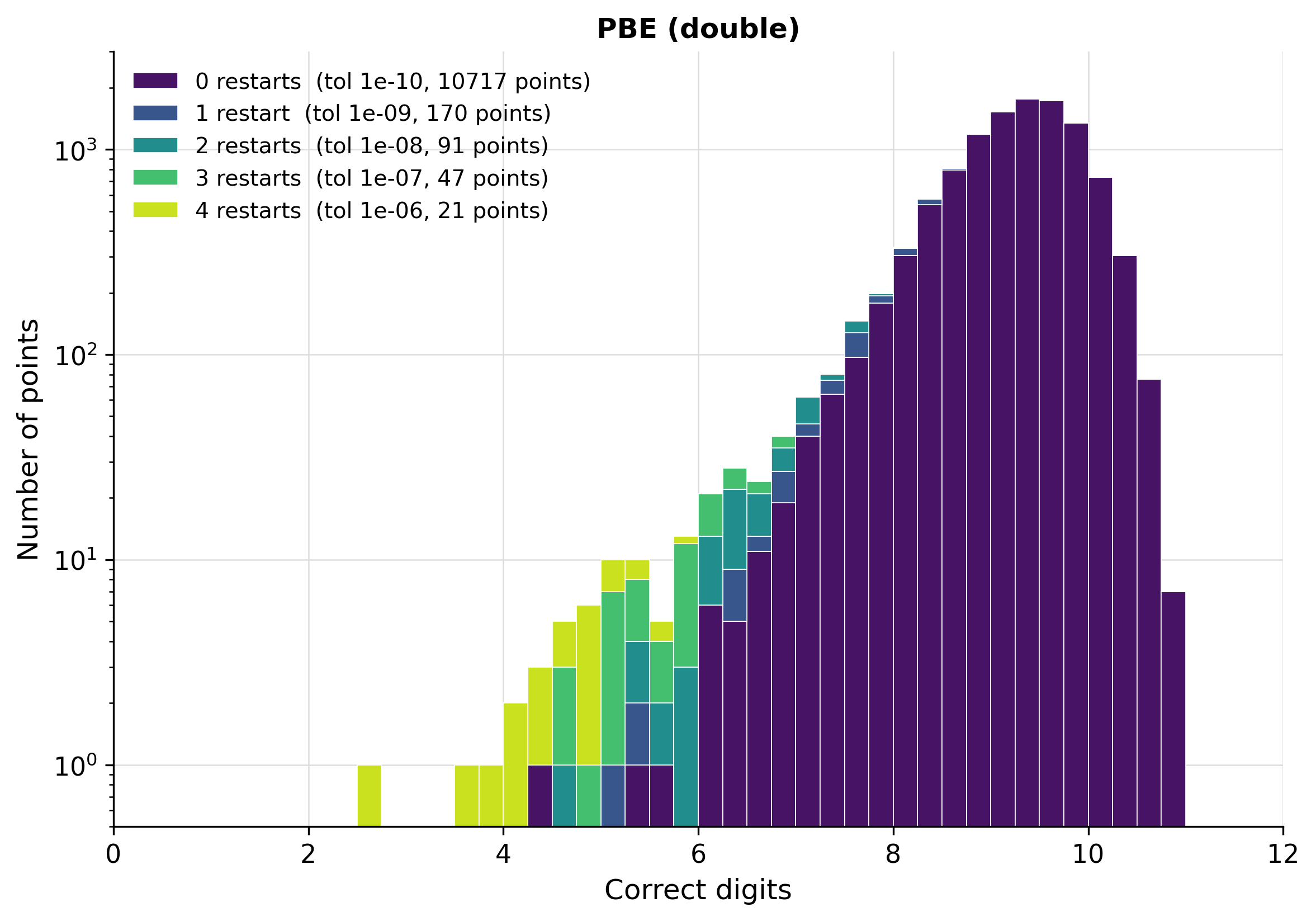}
\caption{Evaluation of the special functions of \PBE\ in double precision. 
Left: distribution of the evaluation time per point, separately for the straight-line and physical paths. 
The hatched component corresponds to points that required at least one restart.
Right: distribution of the correct digits, with each colour corresponding to a different tolerance.}
\label{fig:PBE}
\end{figure}

\begin{figure}[t]
\centering
\includegraphics[width=0.48\textwidth]{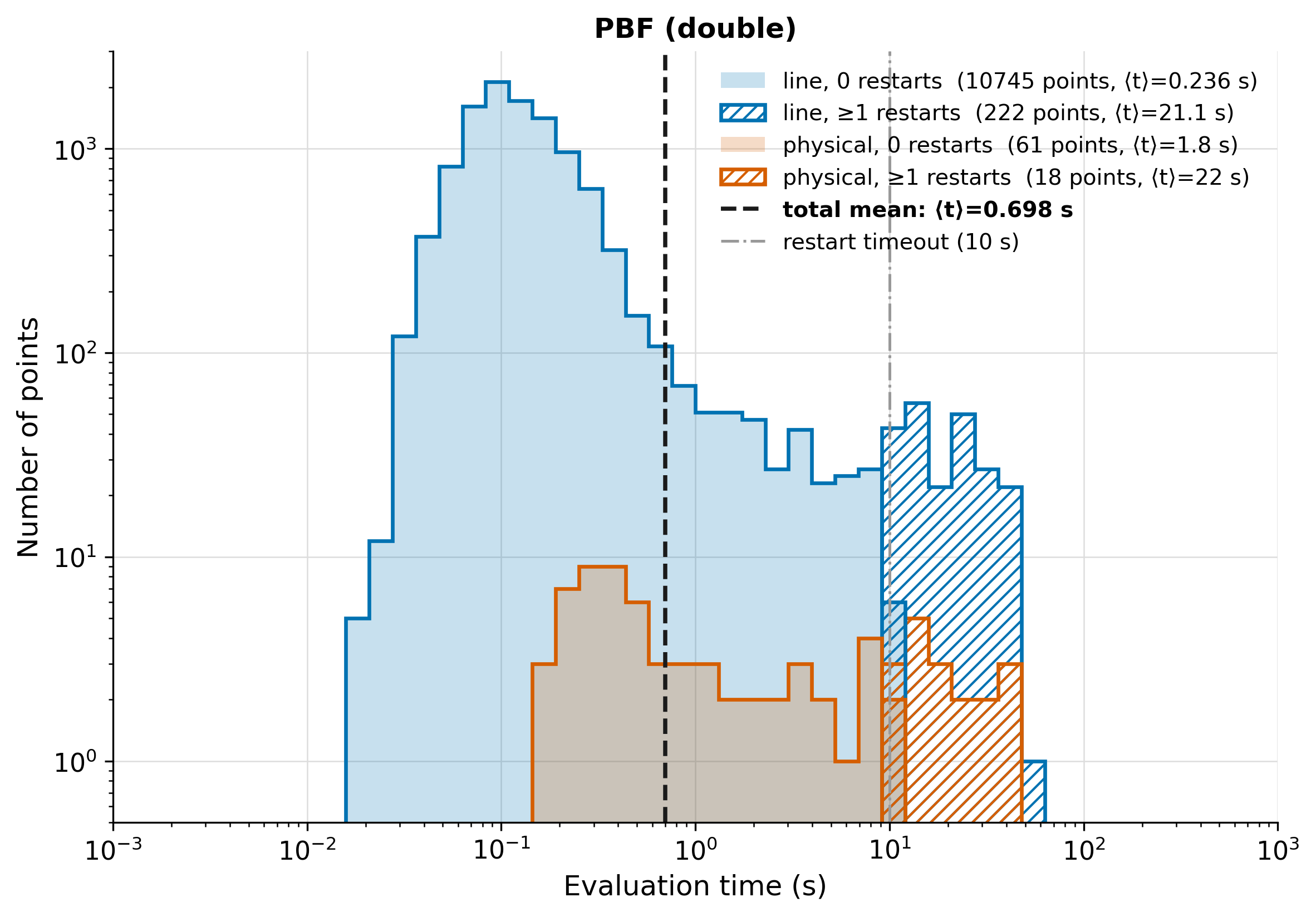}\hfill
\includegraphics[width=0.48\textwidth]{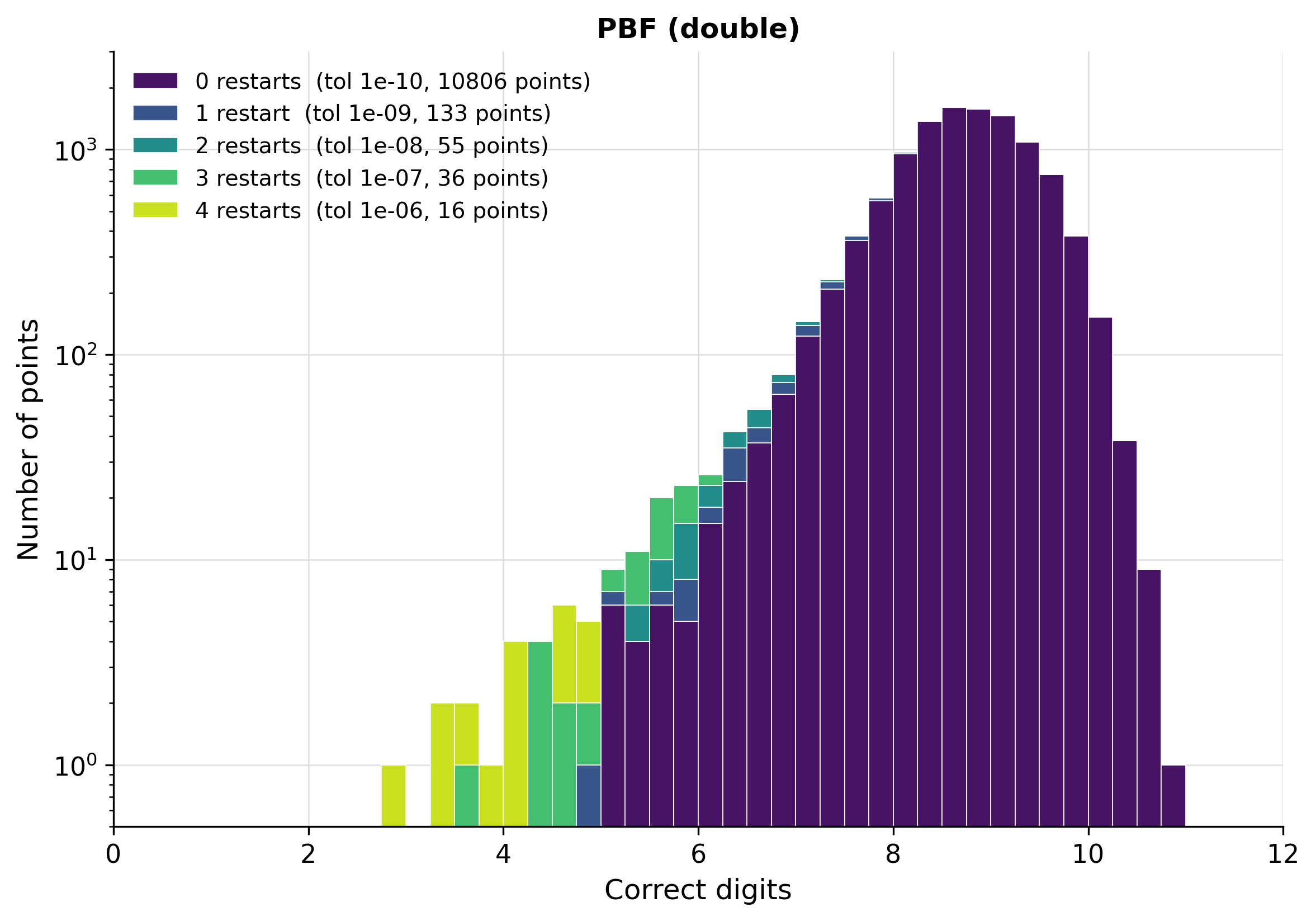}
\caption{Evaluation of the special functions of \PBF\ in double precision. 
Left: distribution of the evaluation time per point, separately for the straight-line and physical paths. 
The hatched component corresponds to points that required at least one restart.
Right: distribution of the correct digits, with each colour corresponding to a different tolerance.}
\label{fig:PBF}
\end{figure}

The results for \PBE\ are shown in \cref{fig:PBE}. 
In this case, the integration becomes slower for a fraction of the points.
We therefore set a time limit of 10\,s.
If the limit is exceeded, we restart the integration with a tolerance larger by a factor of 10, i.e.\ we use successively $\epsilon_{\rm abs}=10^{-10},\dots,10^{-6}$. 
About $96\%$ of the points are evaluated without restarts, with an average
evaluation time of about $0.2$\,s. The remaining points are few, but they
dominate the average, which rises to about $0.8$\,s per point. Only 10 points
could not be evaluated within the time limit even with $\epsilon_{\rm abs}=10^{-6}$. 
We integrated these points in quadruple precision with a tolerance of $10^{-10}$, which took about $4$\,s per point and yielded between 6 and 9 correct digits.
We stress that the impact of these points on the total average evaluation time is negligible.
For this family, the change of one-form basis in \cref{eq:rotation_of} yields a $9\%$ reduction in the mean integration time.
We obtained similar results for \PBF, with an average evaluation time of about $0.7$\,s. 
We show the results in \cref{fig:PBF}. 
In this case, the change of one-form basis in \cref{eq:rotation_of} reduces the mean integration time by 16--19\%, depending on the starting point. Interestingly, the gain is larger than for \PBE, even though the change of basis lowers the maximal polynomial degree more for \PBE{} than for \PBF.
This indicates that the maximal polynomial degree is only a partial measure of the algebraic complexity of the expressions.
Again, 10 points, 6 of which are shared with \PBE, could not be evaluated in double precision within the time limit with $\epsilon_{\rm abs}=10^{-6}$.
Integrating them in quadruple precision with $\epsilon_{\rm abs}=10^{-10}$ took about $10$\,s per point and yielded between 5 and 9 correct digits.
These points were not included in the histograms as they were obtained with a different precision setting, and are therefore not directly comparable with the rest of the sample. It would be interesting to study how these problematic points are distributed in phase space, and whether a tailored set of starting points would alleviate this stability issue.
We leave this to the future computation of the amplitudes, where we will implement both a rescue system to handle such points automatically and the amplitude-level error estimate of ref.~\cite{Badger:2025ljy}.
Nonetheless, the comparison with ref.~\cite{Badger:2025ljy} indicates that our results are already suitable for application in phenomenological studies.

\section{Conclusion}
\label{sec:Conclusions}

We have completed the computation of the two-loop Feynman integrals required for the NNLO QCD corrections to on-shell top--antitop pair production in association with a photon at hadron colliders, in the leading-colour approximation. 
The same integrals also contribute to the subleading-colour NNLO QCD corrections to on-shell top--antitop pair production in association with a jet.

For one of the three new integral families, we constructed a basis of master integrals that satisfies differential equations in canonical form. 
This allowed us to express the solution in terms of a set of algebraically independent special functions, up to the order in $\eps$ required for NNLO QCD applications~\cite{Gehrmann:2018yef,Chicherin:2020oor,Chicherin:2021dyp,Abreu:2023rco}.
The other two families involve multiple elliptic curves and nested square roots. 
Given this higher analytic complexity, we constructed bases for them that satisfy algebraic differential equations in the less constrained form proposed in ref.~\cite{Badger:2024fgb}. 
Most entries of the connection matrices are canonical. 
The remaining entries, which are associated with the elliptic curves and nested square roots, are polynomials in $\eps$ of at most quadratic degree. 
We also chose the master integrals involving elliptic curves and nested square roots so that these analytic structures manifestly contribute only to the $\mathcal{O}\left(\eps^0\right)$ part of the two-loop amplitude.
Following ref.~\cite{Badger:2024dxo}, we expressed the master integrals of both families in terms of special functions. 
The price of the non-canonical form of the differential equations is that a small fraction of these functions may obey relations we have not detected. 
Those functions, however, are precisely the ones associated with the elliptic curves and nested square roots, and they enter the two-loop amplitude only at $\mathcal{O}\left(\eps^0\right)$. 
Any missed relations therefore cannot spoil the exact cancellation of the UV and IR poles.

Our use of algebraic differential equations for the elliptic sectors was a pragmatic choice, implemented heuristically.
Recent studies have begun to investigate whether, in the presence of elliptic curves, there exist preferred bases obeying algebraic or rational differential equations, and have revealed unexpected underlying structures~\cite{Chaubey:2025adn,Chaubey:2026egr}.
It would be interesting to compare the bases constructed here with those suggested by these studies, and to test whether either choice offers advantages for the numerical evaluation or yields simpler expressions for the amplitudes.

Following ref.~\cite{Badger:2025ljy}, we evaluate the special functions numerically for each family by solving the corresponding non-linear systems of rational first-order differential equations using the Bulirsch--Stoer algorithm as implemented in \textsc{Boost}.
We introduced a new approach to simplifying the one-forms in the differential equations, based on the pole decorrelation of ref.~\cite{DeLaurentis:2023nss}.
The same approach could also speed up the evaluation of the special functions for $t\bar{t}+$jet production of ref.~\cite{Badger:2025ljy}.
We validated our numerical evaluation routine against independent evaluations with \textsc{AMFlow} and \textsc{DiffExp}~\cite{Hidding:2020ytt},
and we tested its performance on a realistic sample of phase-space points.
On this basis, we are confident that our results allow a fast and stable numerical evaluation of all master integrals, suitable for phenomenological applications. All results are provided in the ancillary files~\cite{ancillary}, together with a \textsc{Mathematica} script that uses \textsc{DiffExp} to evaluate the special functions.

Our results open the door to the computation of the two-loop leading-colour amplitudes for $t\bar{t}\gamma$ production at hadron colliders. 
They also take a further step towards the subleading-colour contributions to $t\bar{t}+$jet production. 
The integrals considered here have a considerably richer analytic structure than those required for $t\bar{t}+$jet production at leading colour~\cite{Badger:2024fgb}, with several elliptic curves and nested square roots appearing simultaneously. 
That the approach of refs.~\cite{Badger:2024fgb,Badger:2024dxo,Badger:2025ljy} remains effective in this more demanding setting is encouraging, and hints that it may prove useful for other multi-scale two-loop processes with massive internal propagators.

\acknowledgments

We thank Simon Badger, Heribertus Bayu Hartanto and Tiziano Peraro for useful discussions, and Simon Badger, Heribertus Bayu Hartanto and Mattia Pozzoli for comments on the manuscript.
We are grateful to Simon Badger for collaboration on the numerical solution of differential equations, and to Rene Poncelet for sharing the phase-space points used in our tests.
S.B.~has been supported by an appointment to the JRG Program at the APCTP through the Science and Technology Promotion Fund and Lottery Fund of the Korean Government and by the Korean Local Governments~--~Gyeongsangbuk-do Province and Pohang City. C.B.~acknowledges funding from the Italian Ministry of Universities and Research (MUR) through FARE grant R207777C4R and from the Max Planck-IAS-NTU Center for Particle Physics, Cosmology and Geometry. C.B.~also thanks the Pauli Center for Theoretical Studies and the University of Zurich for hospitality.
This work was supported by the European Research Council (ERC) under the European Union's Horizon Europe research and innovation programme grant agreement 101040760, \textit{High-precision multi-leg Higgs and top physics with finite fields} (ERC Starting Grant \emph{FFHiggsTop}).
S.Z.~was supported by the Swiss National Science Foundation (SNSF) under the Ambizione grant No.~215960. 

\appendix

\section{Master integrals for the five-point sectors}
\label{app:New_Sectors}

In this appendix, we collect the genuinely new master integrals with five-point kinematics, i.e.\ those that remain after modding out by permutations of the external momenta and by products of one-loop integrals. For completeness, the list also includes those already discussed in the main text. Rather than listing their explicit expressions, which can be rather lengthy, we indicate their position within the basis of the corresponding integral family. The full definitions can be found in the ancillary files~\cite{ancillary}. The sectors are classified according to the number of propagators they contain, starting with the six-propagator sectors and ending with the eight-propagator ones.

\raggedbottom

\MISectorGroup{Six-Propagator Sectors}

\MISector{I_{11110011000}^{(\PBD)}}
  {\includegraphics[scale=0.38]{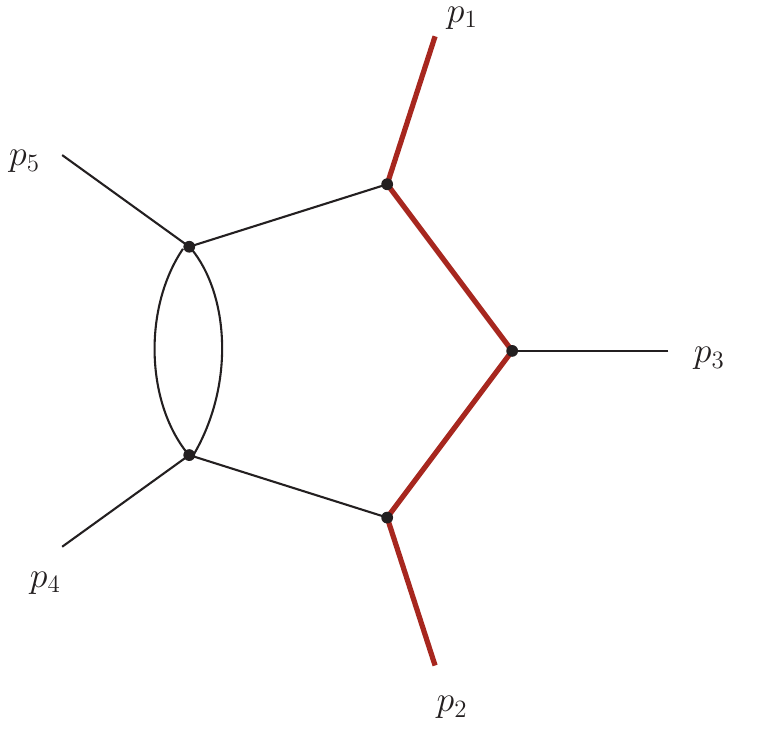}}
  {\mathcal{J}^{(\PBD)}_{26},\mathcal{J}^{(\PBD)}_{27}}

\MISector{I_{11110101000}^{(\PBE)}}
  {\includegraphics[scale=0.38]{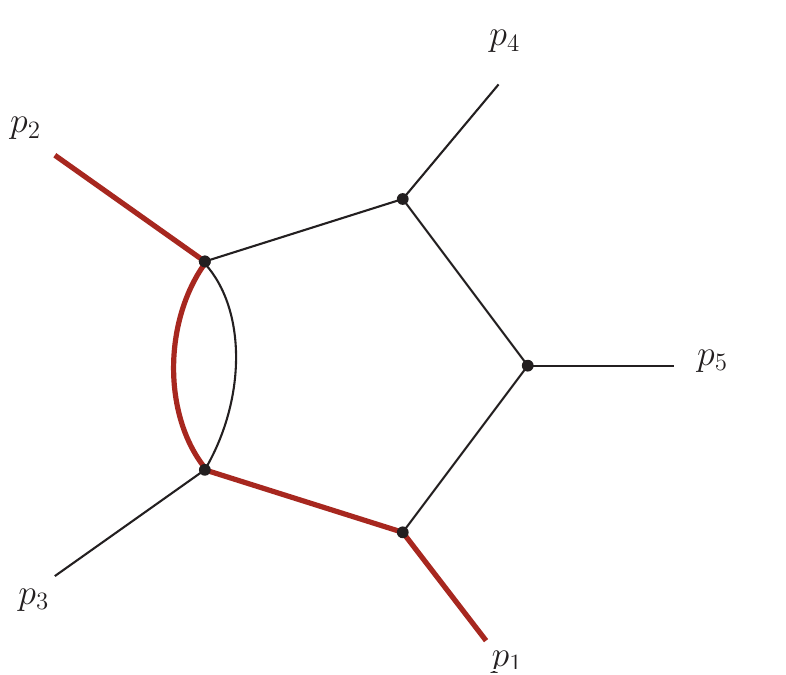}}
  {\mathcal{J}^{(\PBE)}_{45},\mathcal{J}^{(\PBE)}_{46},\mathcal{J}^{(\PBE)}_{47}}

\MISector{I_{11110101000}^{(\PBF)}}
  {\includegraphics[scale=0.38]{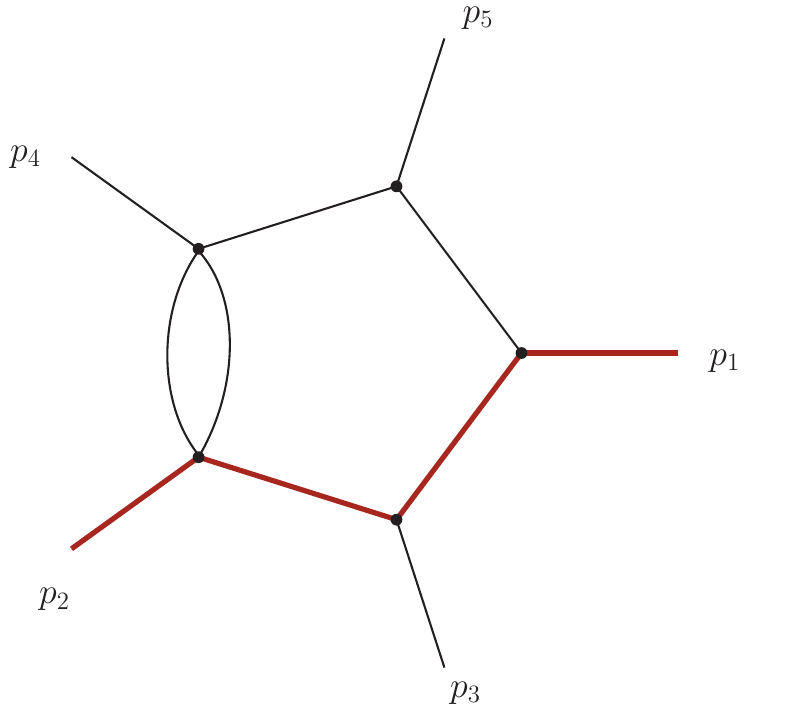}}
  {\mathcal{J}^{(\PBF)}_{49},\mathcal{J}^{(\PBF)}_{50}}

\MISector{I_{01111011000}^{(\PBD)}}
  {\includegraphics[scale=0.52]{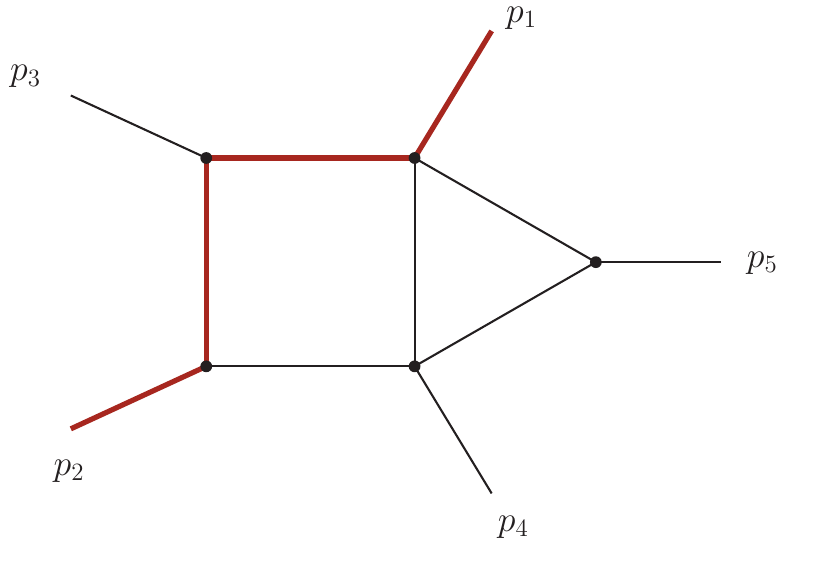}}
  {\mathcal{J}^{(\PBD)}_{22},\mathcal{J}^{(\PBD)}_{23}}

\MISector{I_{01101111000}^{(\PBD)}}
  {\includegraphics[scale=0.52]{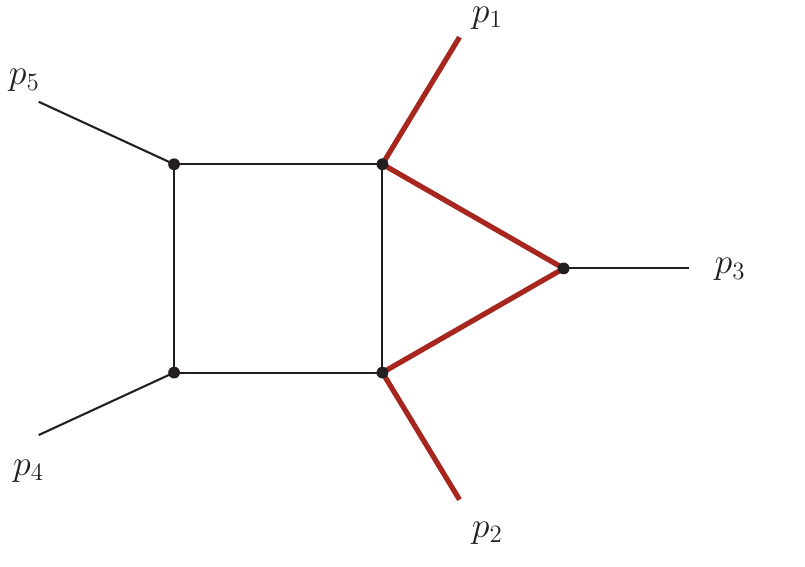}}
  {\mathcal{J}^{(\PBD)}_{18},\mathcal{J}^{(\PBD)}_{19},\mathcal{J}^{(\PBD)}_{20},\mathcal{J}^{(\PBD)}_{21}}

\MISector{I_{11100111000}^{(\PBE)}}
  {\includegraphics[scale=0.52]{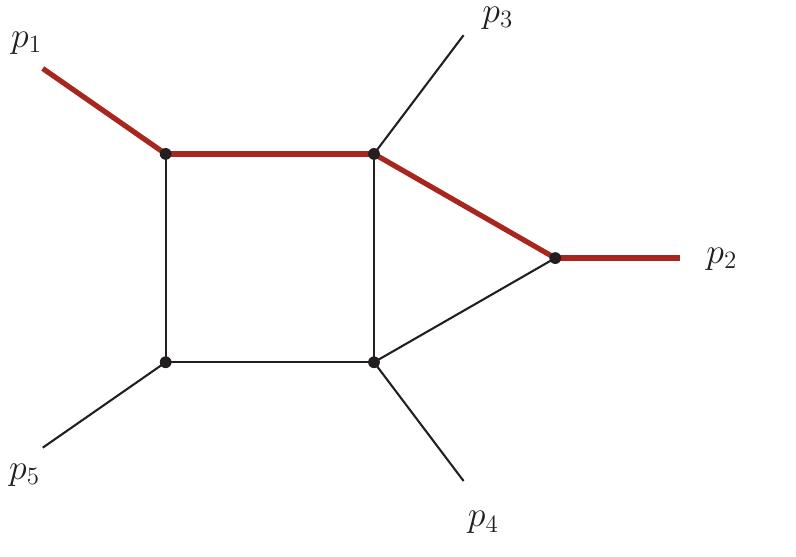}}
  {\mathcal{J}^{(\PBE)}_{26},\mathcal{J}^{(\PBE)}_{27},\mathcal{J}^{(\PBE)}_{28},\mathcal{J}^{(\PBE)}_{29},\mathcal{J}^{(\PBE)}_{30}}

\MISectorGroup{Seven-Propagator Sectors}

\MISector{I_{11111101000}^{(\PBE)}}
  {\includegraphics[scale=0.40]{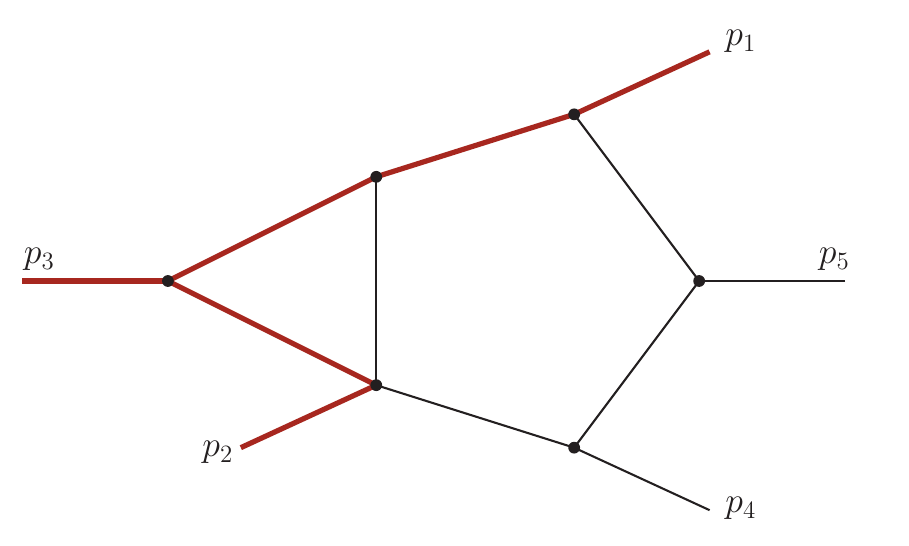}}
  {\mathcal{J}^{(\PBE)}_{23},\mathcal{J}^{(\PBE)}_{24}}

\MISector{I_{11111101000}^{(\PBF)}}
  {\includegraphics[scale=0.40]{Figs/PBF_nested.pdf}}
  {\mathcal{J}^{(\PBF)}_{17},\mathcal{J}^{(\PBF)}_{18}, \mathcal{J}^{(\PBF)}_{19}}

\MISector{I_{11101111000}^{(\PBD)}}
  {\includegraphics[scale=0.50]{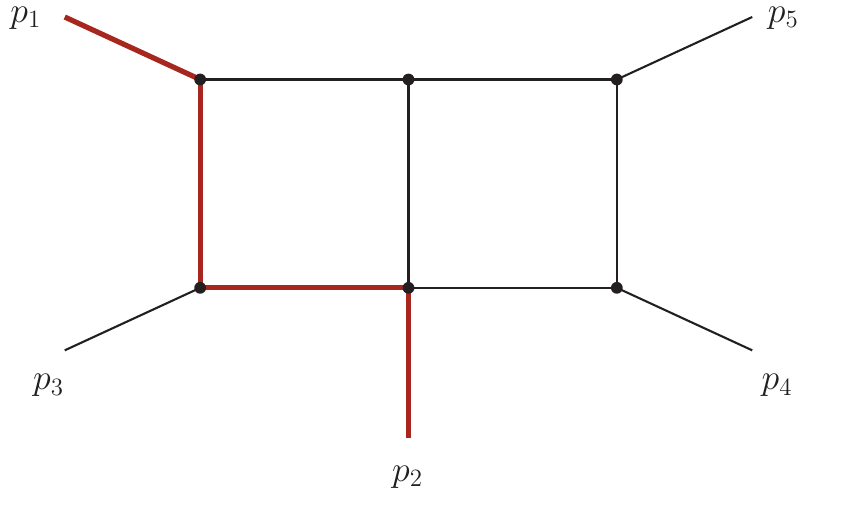}}
  {\mathcal{J}^{(\PBD)}_{8},\mathcal{J}^{(\PBD)}_{9},\mathcal{J}^{(\PBD)}_{10},\mathcal{J}^{(\PBD)}_{11}}

\MISector{I_{11101111000}^{(\PBE)}}
  {\includegraphics[scale=0.50]{Figs/PBE_nested.pdf}}
  {\begin{aligned}
    &\mathcal{J}^{(\PBE)}_{5}, \mathcal{J}^{(\PBE)}_{6}, \mathcal{J}^{(\PBE)}_{7}, \mathcal{J}^{(\PBE)}_{8},\\
    &\mathcal{J}^{(\PBE)}_{9}, \mathcal{J}^{(\PBE)}_{10}, \mathcal{J}^{(\PBE)}_{11}
  \end{aligned}}

\MISector{I_{01111111000}^{(\PBF)}}
  {\includegraphics[scale=0.48]{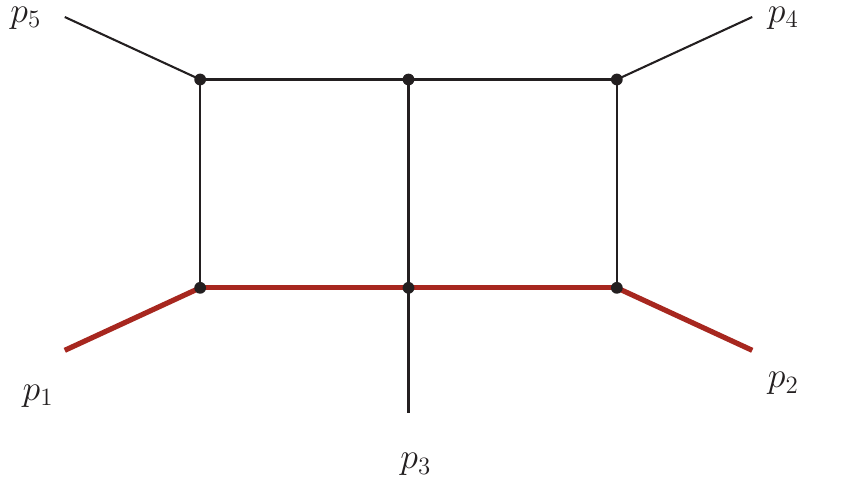}}
  {\mathcal{J}^{(\PBF)}_{12},\mathcal{J}^{(\PBF)}_{13}}

\MISectorGroup{Eight-Propagator Sectors}

\MISector{I_{11111111000}^{(\PBD)}}
  {\includegraphics[scale=0.44]{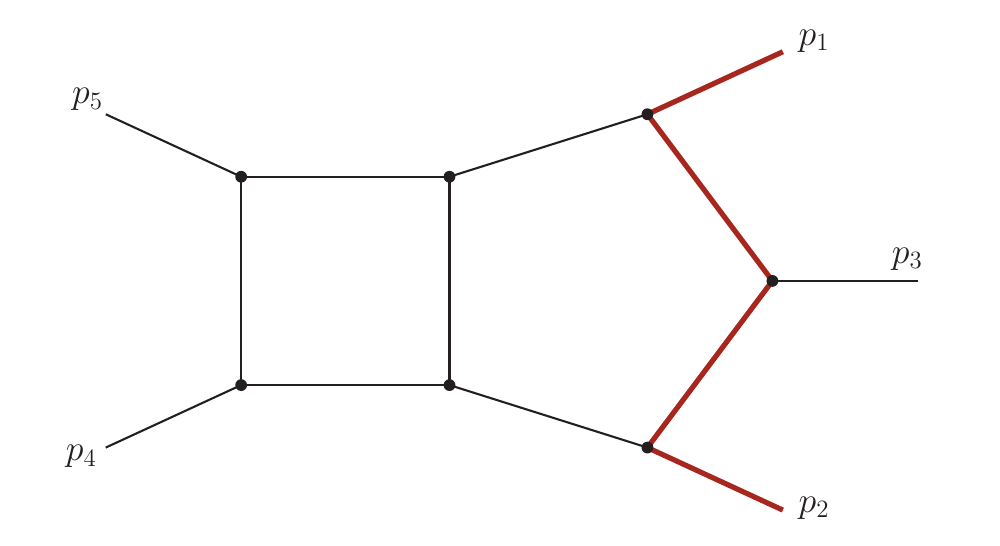}}
  {\mathcal{J}^{(\PBD)}_{1},\mathcal{J}^{(\PBD)}_{2},\mathcal{J}^{(\PBD)}_{3}}

\MISector{I_{11111111000}^{(\PBE)}}
  {\includegraphics[scale=0.44]{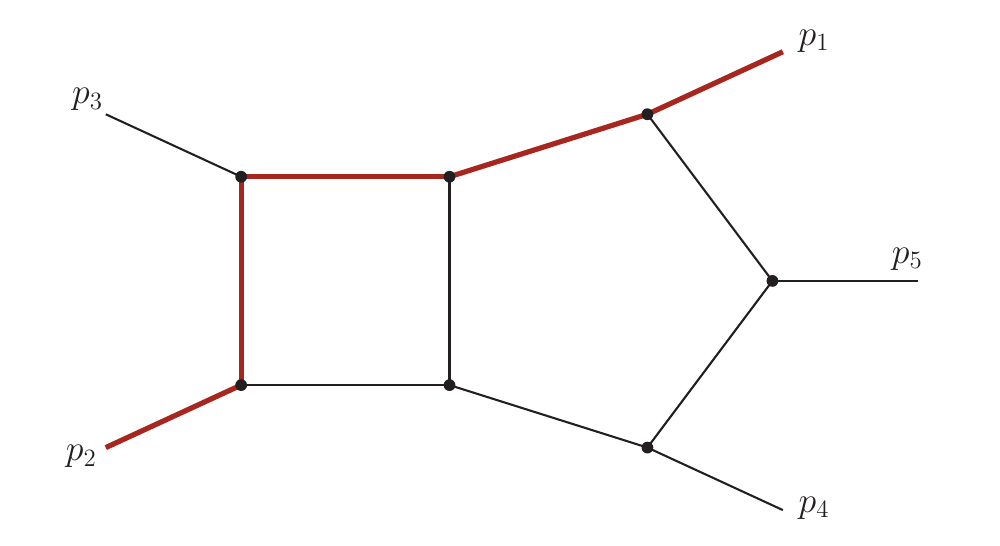}}
  {\mathcal{J}^{(\PBE)}_{1},\mathcal{J}^{(\PBE)}_{2},\mathcal{J}^{(\PBE)}_{3},\mathcal{J}^{(\PBE)}_{4}}

\MISector{I_{11111111000}^{(\PBF)}}
  {\includegraphics[scale=0.44]{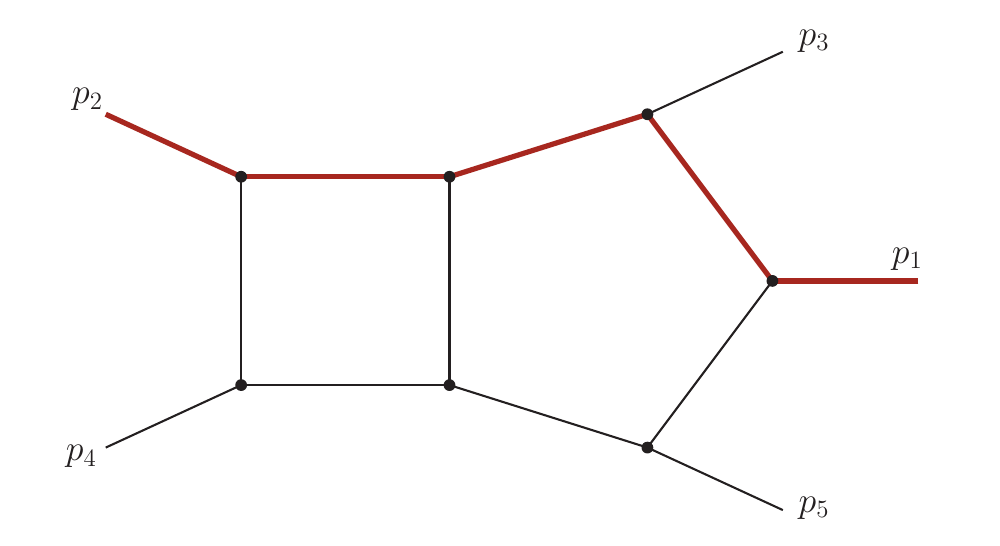}}
  {\mathcal{J}^{(\PBF)}_{1},\mathcal{J}^{(\PBF)}_{2},\mathcal{J}^{(\PBF)}_{3}}

\par\pagebreak[0]
\flushbottom

\bibliographystyle{JHEP}
\bibliography{biblio}

\end{document}